\documentclass[a4paper,fleqn]{cas-dc}

\usepackage[numbers]{natbib}
\usepackage{subcaption}
\usepackage{pdflscape}   
\usepackage{array}       
\usepackage{booktabs}    
\usepackage{graphicx} %
\usepackage{tabularx}   
\usepackage{multirow}   
\usepackage{caption}    
\usepackage{svg} 
\usepackage{stfloats}
\usepackage{float}
\usepackage[acronym]{glossaries}
\usepackage{amsmath,amssymb,booktabs,longtable,array}
\usepackage[T1]{fontenc}
\usepackage[utf8]{inputenc}
\usepackage{xcolor}
\usepackage{hyperref}
\usepackage{cuted}
\usepackage{algorithm}
\usepackage{algpseudocode}

\makeglossaries

\newacronym{ems}{EMS}{Energy Management System}
\newacronym{bems}{BEMS}{Battery Energy Management System}
\newacronym{bess}{BESS}{Battery Energy Storage System}
\newacronym{soc}{SOC}{State of Charge}
\newacronym{soh}{SOH}{State of Health}
\newacronym{ocv}{OCV}{Open-Circuit Voltage}
\newacronym{rl}{RL}{Reinforcement Learning}
\newacronym{tfe}{TFE}{Temporal Feature Extractor}
\newacronym{lstm}{LSTM}{Long Short-Term Memory}
\newacronym{mse}{MSE}{Mean Square Error}
\newacronym{mlp}{MLP}{Multilayer Perceptron}
\newacronym{mae}{MAE}{Mean Absolute Error}
\newacronym{der}{DER}{Distributed Energy Resources}
\newacronym{mpc}{MPC}{Model Predictive Control}
\newacronym{drl}{DRL}{Deep Reinforcement Learning}
\newacronym{il}{IL}{Imitation Learning}
\newacronym{bc}{BC}{Behavior Cloning}
\newacronym{dagger}{DAgger}{Dataset Aggregation}
\newacronym{pv}{PV}{Photovoltaic}
\newacronym{lp}{LP}{Linear Programming}
\newacronym{lfp}{LFP}{Lithium Iron Phosphate}
\newacronym{lut}{LUT}{Look-up Table}
\newacronym{ppo}{PPO}{Proximal Policy Optimization}
\newacronym{nlp}{NLP}{Non-Linear Programming}
\newacronym{milp}{MILP}{Mixed-Integer Linear Programming}
\newacronym{gail}{GAIL}{Generative Adversarial Imitation Learning}
\newacronym{airl}{AIRL}{Adversarial Inverse Reinforcement Learning}
\newacronym{dp}{DP}{Dynamic Programming}
\newacronym{tou}{ToU}{Time of Use}
\newacronym{dam}{DAM}{Day-Ahead Market}
\newacronym{sac}{SAC}{Soft Actor–Critic}
\newacronym{dqn}{DQN}{Deep Q-Network}
\newacronym{r2d}{R2D}{Representation-to-Decision}
\newacronym{tcn}{TCN}{Temporal Convolutional Network}
\newacronym{llm}{LLM}{large language model}
\newacronym{ml}{ML}{machine learning}
\newacronym{ood}{OOD}{Out-of-Distribution}
\newacronym{pca}{PCA}{Principal Component Analysis}
\def\tsc#1{\csdef{#1}{\textsc{\lowercase{#1}}\xspace}}
\tsc{WGM}
\tsc{QE}

\begin{document}
\let\WriteBookmarks\relax
\def\floatpagepagefraction{1}
\def\textpagefraction{.001}

\shorttitle{Learning Control Policies from Heterogeneous Time Series in BEMS}    

\shortauthors{Yin et al.}  

\title [mode = title]{Learning Control Policies from Heterogeneous Multi-Horizon Time Series in Battery Energy Management Systems}  


%

\author[1,2]{Sheng Yin}[orcid=0009-0000-6059-0047]
\cormark[1]

\credit{Conceptualization, Methodology, Software, Data curation, Writing – original draft, Writing – review \& editing, Formal analysis, Validation, Visualization, Project administration}
\ead{sheng.yin@tum.de}

\author[1,2]{Vivek Teja Tanjavooru}[orcid=0000-0001-9042-8067]
\ead{vivek.tanjavooru@tum.de}
\credit{Software, Validation, Writing – review \& editing}

\author[2]{Holger Hesse}[orcid=0000-0002-6670-2684]
\ead{holger.hesse@hs-kempten.de}
\credit{Supervision, Writing – review \& editing, Funding acquisition}

\author[1]{Christoph Goebel}[orcid=0000-0002-5756-6983]
\ead{christoph.goebel@tum.de}
\credit{Supervision, Writing – review \& editing}

\affiliation[1]{organization={Technical University of Munich, TUM School of Engineering and Design, Department of Energy and Process Engineering, Chair of Energy Management Technologies},
            addressline={Arcisstraße 21}, 
            city={Munich},
            postcode={80333}, 
            country={Germany}}
            
\affiliation[2]{organization={Kempten University of Applied Sciences, Institute for Energy and Propulsion Technologies},
            addressline={Bahnhofstr. 61}, 
            city={Kempten},
            postcode={87435}, 
            country={Germany}}
\cortext[1]{Corresponding author at: Technical University of Munich, TUM School of Engineering and Design, Department of Energy and Process Engineering, Chair of
Energy Management Technologies, Arcisstr. 21, 80333 Munich, Germany.}


\begin{abstract}
This paper introduces Representation-to-Decision (R2D), an end-to-end imitation learning framework that maps heterogeneous multi-horizon time-series inputs directly to battery control decisions through modular Temporal Feature Extractors (TFEs) and a shared latent representation, without an explicit load and PV forecasting step.
While accurate forecasting improves prediction quality, optimal control performance remains unguaranteed in prediction-then-optimization pipelines; standard Reinforcement Learning (RL) lacks the long-horizon temporal awareness due to short-window observations, even when forecast signals are available as additional inputs.
R2D offers a different perspective: rather than forecasting first and deciding second, it learns to decide directly from raw temporal inputs, with control optimality anchored by an aging-aware Mixed-Integer Linear Programming (MILP) expert through Behavior Cloning (BC).
Benchmarked against six controllers on a high-fidelity electro-thermal battery simulation across five industrial sites, R2D achieves 62--77\% of the global clairvoyant optimum, outperforms the tested Model Predictive Control (MPC) and RL benchmarks, and yields battery degradation nearly identical to its MILP teacher across all five sites.
Comprehensive ablation studies over temporal backbone, model size, expert formulation, aging-cost weighting, and horizon configuration confirm that an LSTM encoder with a 15-minute single-step control horizon provides the most robust and deployment-ready configuration, and cross-site and single-factor out-of-distribution tests show that generalization to unseen profiles is profile-dependent, with policies trained on high-activity sites transferring most reliably.
\end{abstract}

\begin{graphicalabstract}
\includegraphics[width=1\columnwidth]{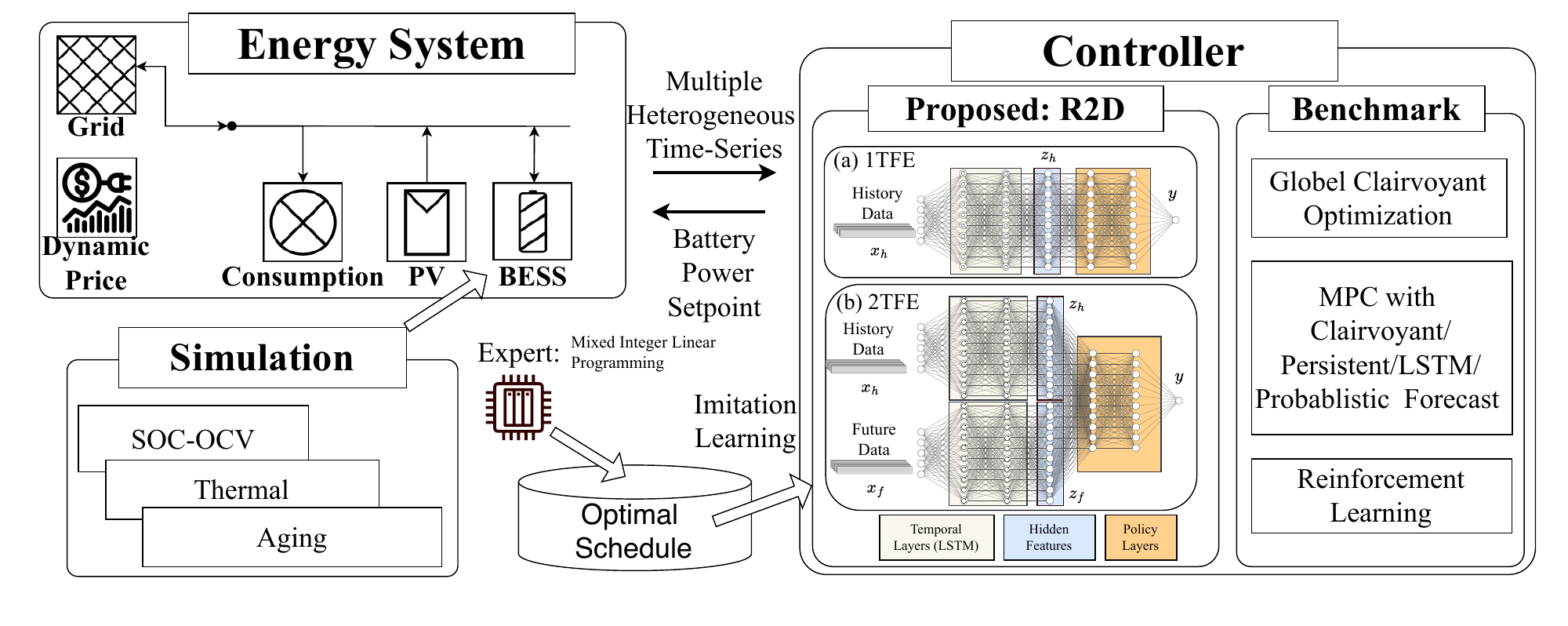}
\end{graphicalabstract}

\begin{highlights}
\item Proposes R2D, an end-to-end imitation learning framework for battery EMS.
\item Encodes load, PV, and day-ahead prices into an implicit temporal representation.
\item Clones an aging-aware MILP expert, inheriting battery-conservative dispatch behavior.
\item Achieves 62--77\% of the global optimum, exceeding the tested MPC and RL baselines.
\item Ablation covers architecture, IL expert, horizon, and cross-site generalization.
\end{highlights}


\maketitle
\newpage
\section{Introduction}
\label{sec:intro}
In the transition toward sustainable energy, \gls{bems} play a key role in integrating renewable energy sources, increasing on-site self-consumption, and mitigating the impact of increasingly volatile electricity tariffs \cite{Luthander2015}. Their operation is inherently affected by multiple sources of uncertainty: electricity demand varies over time, \gls{pv} generation depends on weather conditions, and electricity prices fluctuate throughout the day, posing significant challenges for forecasting, planning, and real-time control~\cite{Hesse2017}. Compounding this, accurate battery operation requires modeling coupled electrical, thermal, and aging dynamics, which are complex nonlinear processes that must be captured to prevent premature degradation while maintaining economic performance.

The value of \gls{bems} is realized through energy arbitrage and peak shaving under dynamic \gls{tou} electricity tariffs and day-ahead price signals. Historical reviews demonstrate that energy-saving effects have grown significantly over the decades, with savings in buildings increasing from roughly 11\% to 16\%~\cite{Lee2016}, yet a large share of efficiency potential in the industrial sector remains unexploited~\cite{Schulze2016}. Unlocking this potential requires the smart \gls{ems} to anticipate future load, \gls{pv} generation, and market signals across multiple horizons, a challenge that demands both temporal awareness and robust handling of uncertainty.

Anticipating these signals is the task of forecasting, which has been studied extensively for both electricity demand and \gls{pv} generation. Systematic reviews show that neural networks are highly effective for short-term load prediction while regression and time-series models remain competitive over longer horizons~\cite{Nti2020,Kuster2017}, and probabilistic models such as \gls{lstm} trained with pinball-loss objectives additionally quantify demand uncertainty~\cite{Wang2019}; comparable surveys document a similarly broad landscape of physical, decomposition, and deep-learning methods for \gls{pv} generation~\cite{Gaboitaolelwe2023}. Architecturally, \gls{lstm} and \gls{tcn} backbones~\cite{Zheng2023} have lately been joined by Transformer- and \gls{llm}-based forecasters that improve price and charging-load accuracy~\cite{Fan2026EPformer,Fan2026LLM}, and jointly forecasting load, \gls{pv}, and price within a single multi-channel encoder outperforms independent models when uncertainty arises from multiple coupled sources~\cite{Li2024}.

Crucially, however, such forecasters are typically trained in isolation from the control task they serve, minimizing statistical error metrics such as root mean square error or continuous ranked probability score rather than operational cost. A statistically superior forecast is therefore not necessarily an economically superior one: asymmetric tariffs and prediction bias shape the true outcome, so a more accurate model can still yield worse dispatch decisions~\cite{Katholnigg2025}. This gap between forecasting accuracy and decision quality motivates examining how the temporal and decision-making components are combined in practical \gls{ems}, which we review next.

\begin{strip}
\noindent\fbox{%
  \begin{minipage}{\dimexpr\textwidth-2\fboxsep-2\fboxrule\relax}
    \small
    \textbf{Nomenclature}\\[4pt]
    %
    \begin{minipage}[t]{0.50\linewidth}
      \textit{Acronyms}\\[2pt]
      \begin{tabular}{@{}l@{\hspace{6pt}}l@{}}
        BC    & Behavior Cloning \\
        BEMS  & Battery Energy Management System \\
        BESS  & Battery Energy Storage System \\
        DAM   & Day-Ahead Market \\
        DQN   & Deep Q-Network \\
        DRL   & Deep Reinforcement Learning \\
        EMS   & Energy Management System \\
        GAIL  & Generative Adversarial Imitation Learning \\
        IL    & Imitation Learning \\
        LFP   & Lithium Iron Phosphate \\
        LP    & Linear Programming \\
        LSTM  & Long Short-Term Memory \\
        LUT   & Look-up Table \\
        MAE   & Mean Absolute Error \\
        MILP  & Mixed-Integer Linear Programming \\
        MLP   & Multilayer Perceptron \\
        MPC   & Model Predictive Control \\
        MSE   & Mean Square Error \\
        OCV   & Open-Circuit Voltage \\
        PPO   & Proximal Policy Optimization \\
        PV    & Photovoltaic \\
        R2D   & Representation-to-Decision \\
        RL    & Reinforcement Learning \\
        SAC   & Soft Actor--Critic \\
        SOC   & State of Charge \\
        SOH   & State of Health \\
        TCN   & Temporal Convolutional Network \\
        TFE   & Temporal Feature Extractor \\
        ToU   & Time of Use \\
      \end{tabular}
      \vspace{5pt}

      \textit{Variables}\\[2pt]
      \begin{tabular}{@{}l@{\hspace{6pt}}p{0.60\linewidth}@{}}
        $t$, $T$                           & Time index; set of all time steps \\
        $s_{t}$, $a_{t}$                   & System state; control action \\
        $\Delta t$                         & Simulation time step (h) \\
        $T_{h}$, $T_{f}$, $T_{c}$         & Historical, forecast, control horizons (h) \\[3pt]
        $p^{G}_{t}$                        & Net grid power (kW) \\
        $p^{G,buy}_{t}$, $p^{G,sell}_{t}$ & Grid import / export power (kW) \\
        $p^{L}_{t}$                        & Site consumption load (kW) \\
        $p^{PV}_{t}$                       & PV generation (kW) \\
      \end{tabular}
    \end{minipage}%
    \hfill
    \begin{minipage}[t]{0.50\linewidth}
      \textit{Variables}\\[2pt]
      \begin{tabular}{@{}l@{\hspace{6pt}}p{0.58\linewidth}@{}}
        $p^{B,AC}_{t}$                         & Battery AC-side power (kW) \\
        $\bar{p}^{B,AC}_{t}$                   & Requested AC-side setpoint (kW) \\
        $p^{B,ch}_{t}$, $p^{B,dch}_{t}$       & Charging / discharging power, AC (kW) \\
        $p^{B,ch,dc}_{t}$, $p^{B,dch,dc}_{t}$ & Charging / discharging power, DC (kW) \\
        $p^{inv,ch}_{t}$, $p^{inv,dch}_{t}$   & Inverter loss, charging /discharging (kW) \\
        $p^{heat}_{t}$                         & Internal heat generation (kW) \\
        $p^{N}$                                & Battery nominal rated power (kW) \\
        $E^{B}_{t}$, $E^{N}$                  & Stored energy; nominal capacity (kWh) \\[3pt]
        $soc^{B}_{t}$                          & State of charge \\
        $soc_{\min}$, $soc_{\max}$             & SOC lower / upper bounds \\
        $soh^{B}_{t}$                          & State of health \\
        $ocv^{B}_{t}$                          & Open-circuit voltage (V) \\
        $i^{B}_{t}$                            & Battery current (A) \\
        $\tau^{B}_{t}$                          & Battery temperature ($^{\circ}$C) \\
        $T_{\mathrm{air}}$                     & Ambient temperature ($^{\circ}$C) \\[3pt]
        $\rho^{\mathrm{ToU}}_{t}$              & Time-of-use purchase tariff (€/kWh) \\
        $\rho^{\mathrm{Sell}}$, $\rho^{B}$    & Feed-in tariff; battery investment cost (€/kWh) \\[3pt]
        $c^{E}$, $\mathbb{C}^{E}$,               & Electricity cost, step / total (€) \\
        $\mathbb{C}^{B,cal}$, $\mathbb{C}^{B,cyc}$ & Calendar, cyclic aging costs (€) \\
        $c^{\mathrm{base}}_{cal}$, $c^{T}_{cal}$, $c^{soc}_{cal}$ & Calendar aging cost coefficients \\
        $c^{ch}_{cyc}$, $c^{dch}_{cyc}$       & Cyclic aging cost coefficients (€/kWh) \\
        $J$                                     & Objective function \\
        $u^{B,ch}_{t}$, $u^{B,dch}_{t}$       & Battery charge / discharge binaries \\
        $u^{G,buy}_{t}$, $u^{G,sell}_{t}$     & Grid import / export binaries \\
        $b^{\mathrm{inv}}_{t}$                 & Inverter on/off binary \\
        $M^{\mathrm{grid}}$, $M^{\mathrm{inv}}$, $\epsilon^{\mathrm{inv}}$ & Big-M constants; inverter tolerance (kW) \\
        $k_{1}$, $k_{2}$, $\gamma^{ch}$, $\gamma^{dch}$ & Thermal and DC loss coefficients \\[3pt]
        $\mathbf{x}_{h,t}$, $\mathbf{x}_{f,t}$ & Historical / future input features \\
        $\mathbf{y}_{t}$, $\hat{\mathbf{y}}_{t}$ & R2D output; expert label \\
        $\mathbf{z}^{h}_{t}$, $\mathbf{z}^{f}_{t}$ & Historical / future latent embeddings \\
        $I^{hod}$, $I^{dow}$, $I^{doy}$, $I^{\mathrm{Weekend}}$ & Temporal index embeddings \\
        $\pi_{\theta}$, $\theta$, $\mathcal{L}_{\mathrm{BC}}$ & R2D policy; parameters; BC loss \\[3pt]
        $S_{M}$, $\bar{S}^{B}_{M}$            & Absolute saving; normalized saving \\
        $\Delta soh_{M}$, $\eta$               & SOH loss  round-trip efficiency \\
      \end{tabular}
    \end{minipage}%
  \end{minipage}%
}%
\vspace{6pt}
\end{strip}

\subsection{Related Work}
\label{sec:RW}

\begin{figure*}[t]
    \centering
    \includegraphics[width=1.8\columnwidth]{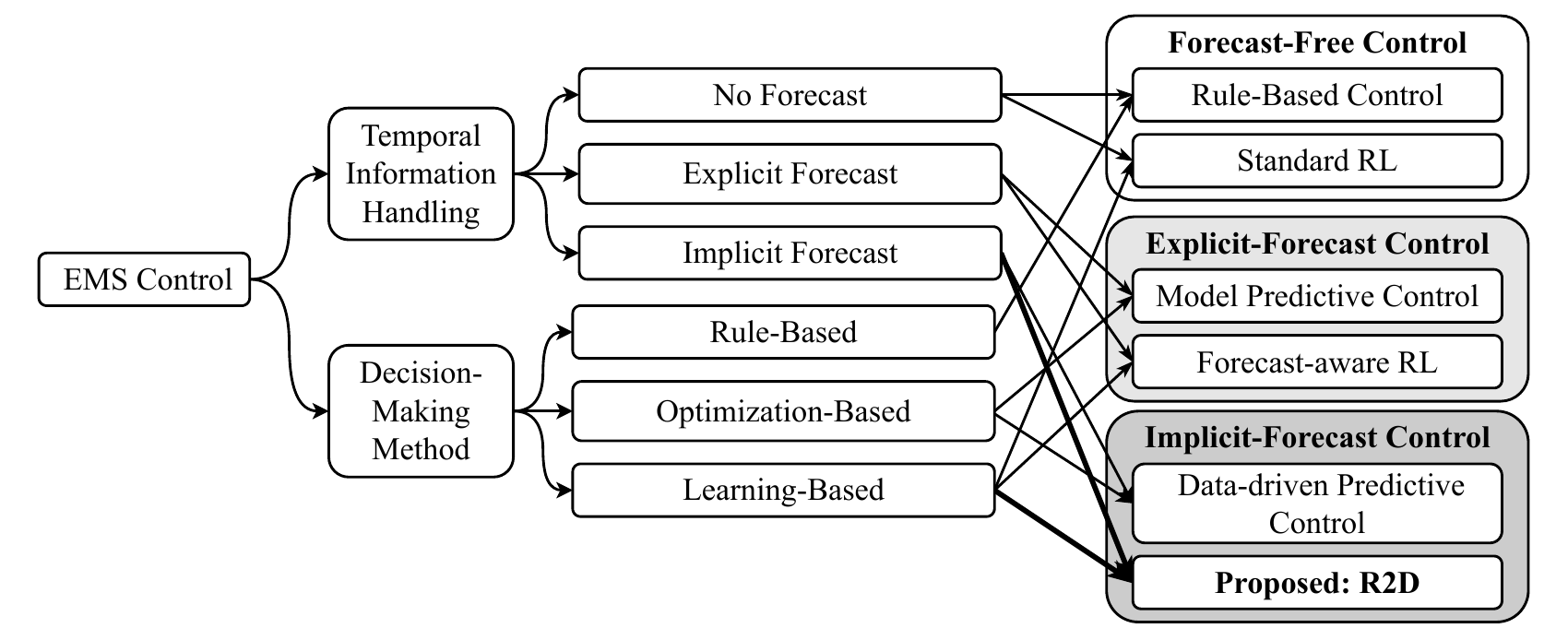}
    \caption{Taxonomy of EMS control approaches along two axes, temporal information handling (no, explicit, or implicit forecast) and decision-making method (rule-based, optimization-based, or learning-based), yielding three control families: forecast-free, explicit-forecast, and implicit-forecast control, with the proposed R2D in the last.}
    \label{fig:taxonomy}
\end{figure*}

We organize existing \gls{ems} control approaches along two axes, temporal information handling and decision-making method, as shown in Figure~\ref{fig:taxonomy}, yielding three control families that differ in how, and whether, future information enters the control loop. Forecast free control makes reactive decisions without any forecast, using rule-based or model-free policies such as rule-based control and standard reinforcement learning. Explicit forecast control achieves temporal awareness through explicit forecasts fed into optimization-based or learning-based controllers, such as \gls{mpc} and forecast-aware reinforcement learning. Implicit forecast control couples a latent forecast representation with decision-making in a single end-to-end architecture, where the proposed \gls{r2d} is positioned.

\subsubsection{Decision-making without Forecast}
Decision-making for \gls{ems} can be broadly divided into three categories: heuristic rule-based (including fuzzy) control, mathematical optimization, and learning-based methods, such as \gls{rl}~\cite{Mischos2023}.

In industrial and residential settings, rule-based control remains the dominant approach due to its reliability, interpretability, and negligible computational cost. Fuzzy-logic controllers extend this approach by encoding expert knowledge into smooth membership functions, which improves the handling of uncertainty in generation, load, and price compared with fixed if--then logic~\cite{Dimitroulis2022,Habib2023}; reported fuzzy \gls{bems} smooth charge--discharge cycles, improve state-of-charge regulation, and extend battery lifetime while remaining suitable for real-time operation~\cite{Habib2023}. However, rule-based control still lacks the flexibility to maximize profits in volatile markets~\cite{Drgo2020}. These heuristic policies are straightforward to deploy and interpret, but their fixed decision rules cannot adapt to the temporal patterns of load, generation, and price, and therefore leave substantial economic value unrealized.

Decision-making for \gls{bems} charging schedules is more commonly formulated as a model-based optimization problem, solved with methods such as genetic algorithms, \gls{lp}, \gls{milp}, or nonlinear programming over a finite horizon. In practical \gls{ems} control, this optimization is embedded in a rolling-horizon, or model predictive control, scheme that relies on forecasts of consumption, \gls{pv} generation, and electricity prices. Collath et al.~\cite{Collath2023} present an \gls{mpc} framework with linearized battery degradation models to improve aging-aware operation, while Kumtepeli et al.~\cite{Kumtepeli2020} develop a three-dimensional mixed-integer linear programming model that captures electro-thermal dynamics and semi-empirical aging effects for arbitrage optimization. Tanjavooru et al.~\cite{Tanjavooru.2026} incorporate spatial thermal dynamics of the battery packs and temperature-dependent thermal derating within a multi-string \gls{mpc} framework to enhance operational safety and efficiency. These models demonstrate the strengths of physics-based formulations; however, they typically address uncertainty only through scenarios or conservative margins, and therefore depend heavily on the quality of the forecast. In practice, perfect forecasts are not achievable because load, \gls{pv} generation, and price signals are highly variable. As a result, \gls{mpc} performance is often limited by forecast errors and the computational effort required for repeated re-optimization.

Learning-based methods provide an alternative to optim-ization-driven control by deriving policies directly from data and environmental interaction. These approaches do not require a detailed analytical model and instead rely on data-centric representations of system behavior, enabling them to operate effectively even when dynamics or constraints are complex or partially unknown. Reinforcement learning and supervised learning are the dominant training strategies~\cite{Mischos2023}. Within this landscape, learning-based controllers are positioned as flexible, model-free components that can be embedded into modern \gls{ems} architectures, complementing sensing and actuation modules while supporting both direct and indirect control schemes.

Most learning-based \gls{ems} methods follow an end-to-end formulation in which control actions are derived solely from the current sensor measurements. As noted in~\cite{Michailidis2025}, many reinforcement learning applications in building and renewable-energy control rely on instantaneous observations rather than structured historical inputs, which restricts the agent's internal representation of system dynamics. Closely related work has shown that seeding \gls{sac} with demonstrations from rule-based policies substantially accelerates convergence on yearly arbitrage tasks~\cite{Sage2023}, confirming that even low-fidelity expert signals provide useful inductive bias. Similarly, Lei et al.~\cite{Lei2025sim} develop a sim-to-real toolchain for RL-based energy management using the distributional \gls{sac} algorithm (DSAC), validating the learned policy through model-in-the-loop and hardware-in-the-loop stages, yet the controller still relies on instantaneous state observations without structured temporal context.

A simple remedy for this lack of temporal context is frame stacking, which has been widely adopted in reinforcement learning research~\cite{arulkumaran2017deep,mnih2015human}. In \gls{ems} applications, stacking two or four observations provides measurable yet modest improvements. According to~\cite{sage2025deep}, stacking four past states increases \gls{dqn} performance by approximately 15\% on residential load--\gls{pv} control tasks and improves \gls{sac} by up to 22\% in energy arbitrage scenarios. However, the benefit of stacking saturates quickly. The same study reports that stacking more than six observations begins to degrade \gls{sac} performance, while \gls{dqn} achieves its best results only with very large stacks of up to 24 frames, which still capture only short-lived patterns rather than long-range temporal dependencies. As a result, although frame stacking introduces limited memory into end-to-end controllers, it remains constrained by small effective horizons and cannot provide the extended temporal structure that \gls{ems} control fundamentally requires.

Imitation learning trains a policy to reproduce expert demonstrations rather than to maximize a reward. In its simplest form, \gls{bc} imitates expert state action pairs~\cite{Mischos2023}, while more advanced variants such as \gls{dagger}, \gls{gail}, and \gls{airl} further address compounding errors or recover implicit rewards. Beyond standard reinforcement learning, imitation learning is also commonly applied in \gls{ems}, where optimization solvers act as the expert: an \gls{milp} solver can label historical data offline and a deep network then imitates its schedule without any online forecast~\cite{Dinh2022}, and \gls{gail} style adversarial imitation has been evaluated for building control~\cite{Liu2024}.

In \gls{ems}, \gls{lp} and \gls{milp} solvers serve as natural experts, consistently outperforming \gls{rl} in cumulative savings under stable conditions~\cite{Yin2025}, and prior work confirms that \gls{bc} from such solutions accelerates convergence and improves performance~\cite{yin2025boosting,Sage2023}. Even when advanced strategies such as imitation learning combined with reinforcement learning are employed~\cite{yin2025boosting}, however, the state representation typically contains only a snapshot of the current load, \gls{pv} generation, price signal, and battery state.

Among these decision-making methods, optimization-based control requires explicit knowledge of the future before the mathematical program can be solved, whereas rule-based control and machine learning can serve as standalone \gls{ems} solutions without any temporal input, reacting directly to the current measurement or observation. This constitutes the forecast-free control family. By relying on snapshot observations without structured temporal context, these approaches operate within a limited observation window that constrains the controller's ability to capture longer-term patterns or anticipate future changes in consumption and renewable production.
\subsubsection{Decision-making with Explicit Forecast}
Based on the standalone forecasting and decision-making reviews, we observe that accurate foresight does not directly translate into good decisions: a good optimization model can provide the perfect solution under perfect information, yet remains highly sensitive to forecast error. Although a forecast-free controller can be developed without forecasting, its long-term performance is limited. A capable \gls{ems} therefore requires temporal awareness and decision-making to be integrated in both design and training, that is, the temporal and decision-making components must be developed and analyzed jointly to obtain a better overall system. Consistent with this view, integrated smart energy management frameworks typically combine monitoring, forecasting, scheduling, and coordination layers to enable intelligent control of \gls{pv}--battery systems~\cite{Zhou2021}.

To overcome the temporal myopia of forecast free control, a second class of methods introduces explicit forecast information into the control loop, forming the explicit forecast control family. Two main variants are found in the literature: \gls{mpc} coupled with a forecasting algorithm, and forecast-aware reinforcement learning.

For optimization-based \gls{mpc}, energy savings between 15\% and 50\% can be achieved~\cite{Drgo2020}, but its practical application is hindered by high modeling effort, the need for specialized expertise, and the inherent decoupling of forecasting and optimization modules~\cite{Drgo2020}. A representative pipeline couples a neural-network (e.g., \gls{lstm}) load and \gls{pv} forecaster with a robust \gls{mpc} optimizer for peak shaving, improving cost performance but making the closed loop directly sensitive to the upstream prediction error~\cite{Mary2025}.

To address the underlying uncertainty, stochastic optimization frameworks for home energy management have been proposed that incorporate scenario-based representations of load and renewable generation variability~\cite{Kim2023}. In both cases, a separately trained forecaster feeds its predictions to a downstream optimization model; the \gls{ems} controller is analyzed as a whole, but the training and development of its two components remain separate.

On the learning-based side, several studies introduce temporal awareness by integrating external forecasts into the state, enriching the agent's observation with predicted future information while preserving a modular architecture. Forecast-aware reinforcement learning methods extend this idea by feeding full forecast trajectories into the policy network, as demonstrated by weather-forecast-aware \gls{dqn}~\cite{shojaeighadikolaei2021weather}, uncertainty-aware \gls{dqn} with Bayesian convolutional neural networks~\cite{lork2020uncertainty}, and forecasting-error-aware \gls{sac}~\cite{li2025forecasting}. More recently, Transformer-based forecasters such as the Temporal Fusion Transformer have been integrated as the prediction stage in hybrid forecast--control pipelines, demonstrating improved multi-horizon accuracy but retaining the modular structure and its associated error propagation risks~\cite{Hu2025}. Although these methods provide stronger temporal context than using only the current state or short frame stacks, they rely on a separately trained forecasting model whose outputs are passed to the \gls{rl} agent. This makes the approach modular rather than end-to-end, and it inherits the typical drawbacks of modular pipelines: information loss between stages, rigid interfaces, and the propagation of forecasting errors directly into the control policy, all of which limit the potential for achieving economically optimal decisions.
\subsubsection{Decision-making with Implicit Forecast}
Across existing \gls{ems} control methods, a common limitation is the modular separation of forecasting and decision-making. Optimization-based controllers explicitly state their objectives and leverage domain knowledge, yet their performance is bound to forecast accuracy, and any prediction error propagates directly into the optimizer. Learning-based modular methods inherit the same issue and additionally lose information when raw data or forecast trajectories are compressed into handcrafted features. Even advanced \gls{rl} approaches that incorporate imitation learning or external forecasts remain modular and cannot jointly adapt prediction and control. End-to-end \gls{rl} avoids this decoupling but typically relies on the current state or short frame stacks, providing only limited temporal awareness and insufficient long-range reasoning. Overall, existing pipelines either discard information through modular interfaces or underutilize the temporal richness available in raw data.

Decision-making with an implicit forecast offers a way out of this trade-off. Although not yet widely explored in \gls{ems}, a few closely related lines of work pursue this idea. Closely related to model-based \gls{mpc}, data-driven predictive control methods aim to bypass explicit system identification while retaining an optimization-based structure. Early work on data-driven simulation and control shows that system responses and control inputs can be constructed directly from measured input--output trajectories without identifying a parametric model, provided sufficient excitation conditions are met~\cite{Markovsky2007}. Building on this behavioral perspective, data-enabled predictive control (DeePC) replaces the model-based prediction step in \gls{mpc} with a non-parametric representation derived from historical data and Hankel matrices, and has been shown to be equivalent to classical \gls{mpc} for linear time-invariant systems~\cite{Coulson2019}. While DeePC removes the explicit system-identification step and preserves hard constraint handling, its guarantees hold only under linear time-invariant dynamics and degrade for the nonlinear electro-thermal and aging behavior of a \gls{bess}~\cite{Coulson2019}.

On the learning-based side, Li et al.~\cite{Li2024TempDRL} develop a \gls{bess} bidding strategy in which a Transformer-based temporal feature extractor learns a latent time-series representation that drives a deep reinforcement learning agent across energy and contingency reserve markets, outperforming both optimization- and \gls{rl}-based benchmarks while remaining more interpretable than black-box policies. This demonstrates the promise of coupling a learned latent representation with decision-making, yet such designs remain largely unexplored for the multi-horizon, aging-aware control of industrial \gls{bess}. R2D departs from this structure by learning a task-oriented latent representation end-to-end, so that the temporal encoding is shaped directly by the control decision rather than fixed in a separate prediction stage.

\subsection{Contribution}
\label{sec:contribution}
The review reveals a persistent gap in \gls{ems} control: in conventional \gls{mpc} the forecast directly supports the decision, but the forecasting model is trained independently of the downstream control objective, minimizing statistical error rather than operational cost, so prediction error propagates into control unweighted by its economic consequence and the temporal richness of raw data is underused. We close this gap with an end-to-end, task-oriented latent representation, offering the \gls{ems} community a transferable design principle, not just a single model:
\begin{itemize}
\item \textbf{An end-to-end learning-based controller deciding directly from time-series inputs.} Modular \gls{tfe}s encode the input time series into implicit temporal representations, on which the control decision is made directly. The controller thereby retains temporal awareness while avoiding the errors introduced by the explicit load and \gls{pv} forecasting stage of a traditional forecast-then-optimize controller. We demonstrate the idea with our implementation \gls{r2d}, whose policy is trained by \gls{bc} from an aging-aware \gls{milp} expert.
\item \textbf{Design guidance for temporal encoding.} Ablations over backbone, model size, expert formulation, aging cost weighting, and horizon identify which design choices matter for deployment, giving reusable guidance rather than one tuned instance.
\item \textbf{Validated performance.} On a high-fidelity electro-thermal \gls{bess} simulation across five industrial sites, R2D in its recommended configuration reaches 62--77\% of the global optimum, outperforms the tested \gls{mpc} and \gls{rl} baselines, reproduces its \gls{milp} teacher's degradation profile, and transfers to unseen profiles under cross-site and single-factor \gls{ood} tests when the training profile is sufficiently informative.
\item \textbf{Open resource.} We release the policy network, \gls{milp} expert, simulation environment, and benchmarks as an open repository~\cite{R2D2026}.
\end{itemize}

\section{Methods}
\label{sec:methods}
This section introduces the grid-connected \gls{pv}--battery use case and its control problem formulation (Section~\ref{sec:energysystem}), the proposed \gls{r2d} framework including its architecture and training procedure (Section~\ref{sec:r2d}), and the experimental design covering the simulation environment, dataset, benchmarks, sensitivity studies and evaluation metrics (Section~\ref{sec:experiments}).

\subsection{Energy System Model}
\label{sec:energysystem}
This section introduces the grid-connected PV--battery energy system operating under a time-of-use (ToU) electricity tariff and its interface with the \gls{ems} controller.

\subsubsection{Use Case}
\label{sec:usecase_description}
\begin{figure*}[t]
    \centering
    \includegraphics[width=2.2\columnwidth]{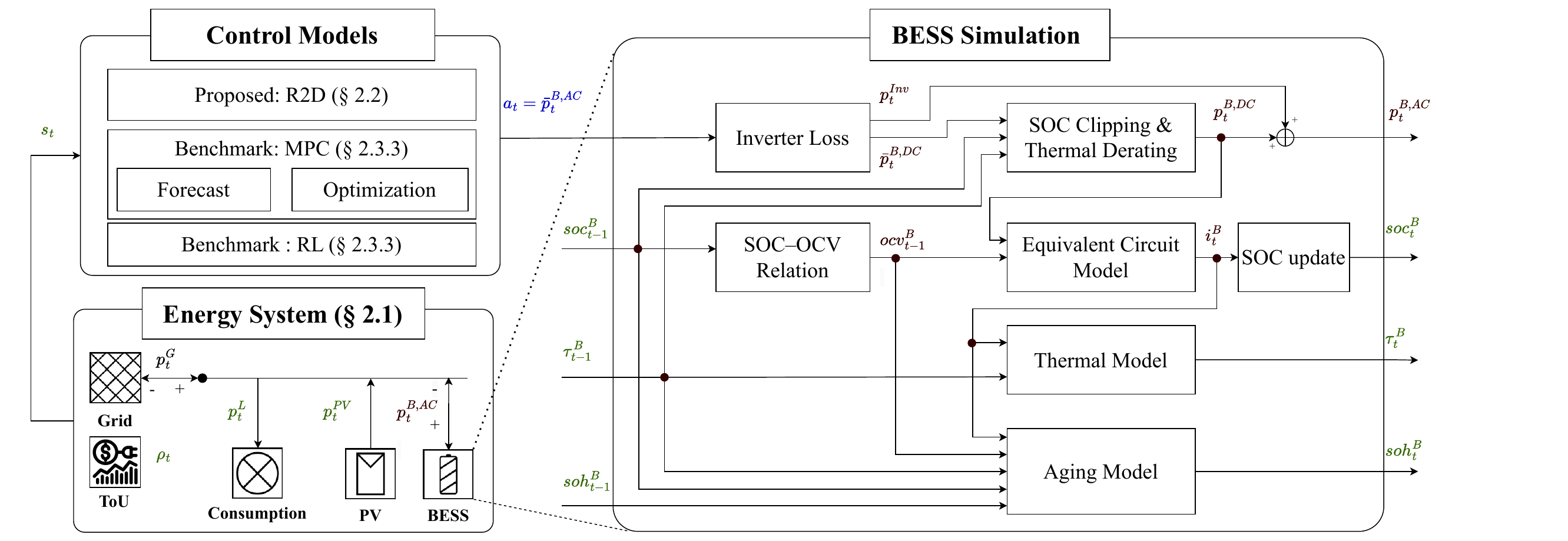}
    \caption{Schematic overview of the use case, including the \gls{ems} simulation with detailed \gls{bess} modeling (Section~\ref{sec:energysystem}) and the proposed \gls{r2d} (Section~\ref{sec:r2d}) and benchmark \gls{ems} control model.}
    \label{fig:usecase}
\end{figure*}

As illustrated in Figure~\ref{fig:usecase}, the industrial site considered in this study is equipped with a \gls{pv} system and a \gls{bess}. The site is connected to the electrical grid and can import or export electricity under a dynamic ToU tariff. The system evolves in discrete time steps of length $\Delta t$.

\subsubsection{Objective}
The \gls{ems} schedules \gls{bess} operation to minimize the total energy cost \(\mathbb{C}^E\), defined as the sum of time-step costs \(c^E_t\):

\begin{equation}
\mathbb{C}^E = \sum_{t \in T} c^E_t= \sum_{t \in T} p^G_t \cdot \Delta t \cdot \rho_t ,
\label{eq:objective_cost}
\end{equation}
where $p^{G}_t$ is the net grid power (positive for import, negative for export) and $\rho_t$ is the corresponding electricity price. The pricing rule is:
\[
\rho_t =
\begin{cases}
\rho^{\mathrm{ToU}}_t, & p^{G}_t \ge 0, \\
\rho^{\mathrm{Sell}},  & p^{G}_t < 0,
\end{cases}
\]
with $\rho^{\mathrm{ToU}}_t$ the time-varying purchase tariff and $\rho^{\mathrm{Sell}}$ the fixed feed-in tariff.

The net grid power follows the system balance:
\begin{equation}
p^{G}_t = p^{L}_t - p^{PV}_t + p^{B,AC}_t ,
\label{eq:grid_balance}
\end{equation}
where $p^{L}_t$ is the site consumption load, $p^{PV}_t$ the \gls{pv} generation, and $p^{B,AC}_t$ the \gls{bess} AC-side power.  

\subsubsection{State Update}
\label{sec:stateupdate}
The consumption load $p_t^L$ and \gls{pv} generation $p_t^{PV}$ are obtained from real industrial site measurements. They inherently contain uncertainty and are treated as time-dependent variables derived directly from the underlying datasets. The ToU purchase price $\rho_t^{\mathrm{ToU}}$ is also time-dependent and obtained from day-ahead electricity market data, representing realistic tariffs available to industrial consumers. The feed-in price $\rho^{\mathrm{Sell}}$ is assumed to be fixed, consistent with typical industrial contract conditions.

The \gls{bess} simulation shown in Fig.~\ref{fig:usecase} estimates the evolution of the battery state variables, \gls{soc} $soc_t^B$, \gls{soh} $soh_t^B$, and temperature $\tau_t^B$. The simulation parameters are based on 60~Ah \gls{lfp} prismatic cells, which determine the electrical and thermal characteristics of the system, as summarized in Table~\ref{tab:para} \cite{Ansean.2013}. At the beginning of each time step ($t$), the inputs for the simulation consist of the optimal power setpoint from the controller $\bar{p}_t^{B,AC}$ and the battery states at the previous time step ($t-1$). These updated states are computed considering the impact of the thermal and degradation dynamics on the battery available capacity. The environment simulates the battery’s internal evolution as:


\begin{equation}
\label{eq:battery_transition}
(soc^B_{t}, soh^B_{t}, \tau^B_{t}) = f_{BESS}(\bar{p}_t^{B,AC}
, soc^B_{t-1}, soh^B_{t-1}, \tau^B_{t-1}).
\end{equation}

This physics-based \gls{bess} simulation framework integrates an experimentally derived inverter loss model, a literature-based SOC--OCV relationship, aging estimation models, and a lumped-mass thermal model to accurately capture the dynamic behavior of the battery under varying operating conditions~\cite{Tanjavooru.2025}. The inverter loss model computes the conversion losses, ${p}_t^{Inv}$, as a function of the requested AC power, $\bar{p}_t^{B,AC}$, yielding the corresponding net electrical power at the DC side, $\bar{p}_t^{B,DC}$. The usable battery power is then constrained through an SOC clipping function that enforces the operable SOC limits. Furthermore, a temperature-dependent power derating factor is applied to account for thermal safety constraints of the \gls{bess}. The resulting net DC power after clipping and derating, ${p}_t^{B,DC}$, combined with the estimated conversion losses, determines the actual AC power exchanged with the grid, ${p}_t^{B,AC}$.

The OCV, ${ocv}_{t-1}^{B}$, is obtained from the SOC--OCV model using the input SOC, ${soc}_{t-1}^{B}$. The battery current, ${i}_{t}^{B}$, is then estimated using a Rint-ECM model that takes as inputs the net DC power, ${p}_{t}^{B,DC}$, and the battery OCV, ${ocv}_{t-1}^{B}$. The estimated current is subsequently used to update the SOC at the end of the time step, ${soc}_{t}^{B}$, via ampere-hour counting. A zero-dimensional thermal model is incorporated to compute the mean battery temperature, $\tau_{t}^{B}$. This model uses the current obtained from the ECM, ${i}_{t}^{B}$, to determine the internal heat generation, and the previous temperature, $\tau_{t-1}^{B}$, together with the cooling parameters, to model heat dissipation. These contributions are combined to update the battery temperature at each simulation step.

Finally, the \gls{soh} is updated at every time step by considering both calendar and cyclic aging, following the methodology described in~\cite{schimpe2018comprehensive}. The SOH model takes as inputs the previous battery states: ${soc}_{t-1}^{B}$, $\tau_{t-1}^{B}$ and ${soh}_{t-1}^{B}$, together with ${ocv}_{t-1}^{B}$, the computed current, ${i}_{t}^{B}$, and the updated SOC, ${soc}_{t}^{B}$. In this work, cyclic aging is further subdivided into three categories: high-temperature cycling, low-temperature cycling, and combined low-temperature and high-\gls{soc} cycling. This provides a detailed and physically consistent representation of battery degradation over the system lifetime.

This combined formulation enables the simulator to capture the coupled electrical, thermal, and aging dynamics of the \gls{bess} with high-fidelity.

\subsubsection{Controller Interface}

\begin{figure}[htbp]
    \centering
    \includegraphics[width=1.\columnwidth]{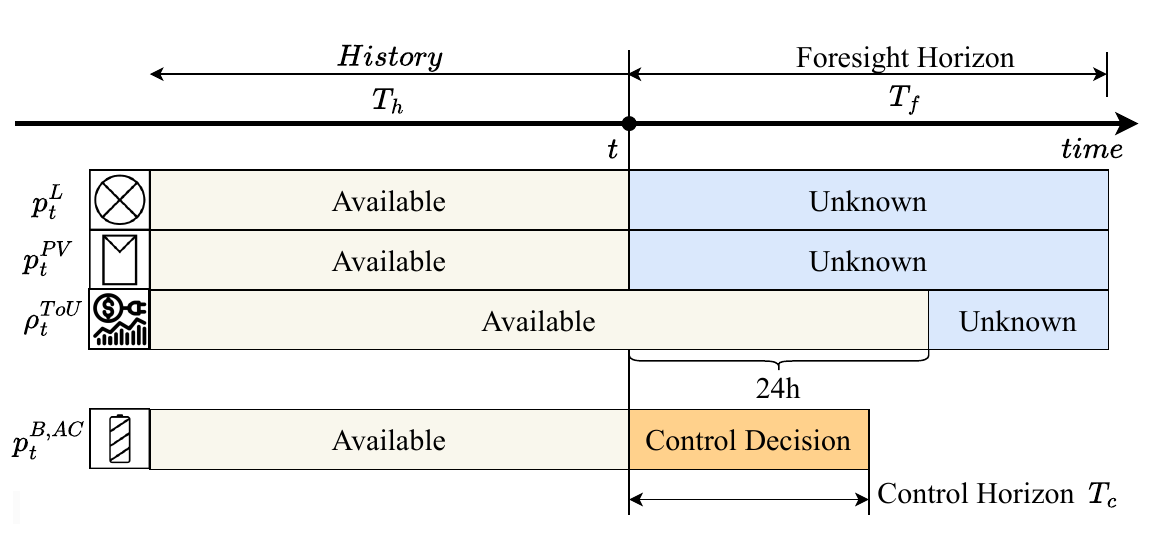}
    \caption{Illustration of the time horizons, showing the information available to the \gls{ems} controller at each decision time, the unknown future variables, and the length of the applied control horizon.
}
    \label{fig:horizon}
\end{figure}

Figure~\ref{fig:horizon} depicts the horizon of raw data available to the \gls{ems} controller and the scope of its decision-making at each time step.

At each time step $t$, the controller has access to all historical measurements of consumption load $p^{L}_{\infty: t - 1}$, \gls{pv} generation $p^{PV}_{\infty: t - 1}$, and battery states. The ToU electricity price is announced one day in advance and is therefore known not only for the past but also for the upcoming 24-hour period, i.e., $\rho^{ToU}_{\infty: t + 24\text{h}}$.

We define the length of usable history as $T_h$, indicating how far back the controller considers past data, while the foresight horizon $T_f$ specifies how far into the future the controller attempts to anticipate system behavior. Based on this information, the controller receives the system state $s_t$ and determines the control action $a_t$. In this framework, the action corresponds to specifying the \gls{bess} AC-side power trajectory $p^{B,AC}_{t : t + T_c}$, which governs battery charging and discharging over the control horizon $[t, t + T_c]$.

\subsection{Proposed Framework: R2D}
\label{sec:r2d}
In this \gls{ems} problem, control complexity arises from the asymmetric temporal availability of information: electricity demand and photovoltaic generation are only observable in the past, whereas electricity prices are typically known over future horizons. To address this challenge, we propose \gls{r2d}, an end-to-end representation-to-decision learning framework that directly maps heterogeneous temporal inputs to control actions. Unlike modular MPC approaches, \gls{r2d} avoids information loss caused by explicit forecast-optimization separation, and unlike conventional learning-based controllers, it provides rich temporal awareness beyond instantaneous or short-window observations. By jointly learning temporal representations and control policies within a unified architecture, \gls{r2d} enables robust decision-making under uncertainty.

In this section, we present the architecture and training schemes of the proposed \gls{r2d} framework.

\subsubsection{Architecture}
\label{sec:architecture}

\begin{figure}[htbp]
\centering
\includegraphics[width=1\columnwidth]{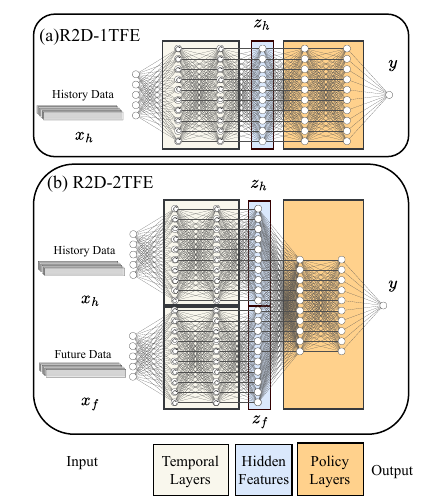}
\caption{Architecture of the proposed \gls{r2d} framework, illustrating the modular temporal feature extractors, hidden feature layers, and policy layers that map raw time-series inputs to control decisions. Two representative instantiations are shown: 
(a)~R2D-1TFE, using a single temporal feature extractor for historical inputs; 
(b)~R2D-2TFE, employing two parallel temporal feature extractors that separately encode historical and future signals.
}
\label{fig:benchmarks_proposed_approach}
\end{figure}

The overall architecture of \gls{r2d} is illustrated in Figure~\ref{fig:benchmarks_proposed_approach}. \gls{r2d} follows a modular representation-to-decision design and consists of three main components: Temporal Feature Extractors \gls{tfe}, hidden feature layers, and policy layers. The \gls{tfe} modules encode heterogeneous temporal inputs into compact latent representations. These representations are then fused in a shared latent space by the hidden layers, after which the policy layers map the fused features to control actions.

Each \gls{tfe} processes sequential input signals, such as load, \gls{pv} generation, or electricity prices, together with time embeddings to capture temporal dependencies over different horizons. The encoded outputs of all instantiated \gls{tfe}s are concatenated and passed through fully connected hidden layers that model cross-temporal and cross-feature interactions. The resulting latent representation is then transformed by the policy layers to generate the control output. This architecture enables \gls{r2d} to jointly learn temporal representations and control policies in an end-to-end manner without an explicit load and PV forecasting module. The model outputs the battery power setpoint for the next control horizon:

\begin{equation}
    \mathbf{y_t} = \bar{p}^{\text{B,AC}}_{t:t+T_c}.
\end{equation}

To accommodate different information structures and time-horizon configurations, we consider two representative instantiations of R2D, summarized in Table~\ref{tab:r2d_variants}, as detailed in the following subsection.

The first variation R2D\_1TFE, shown in Figure~\ref{fig:benchmarks_proposed_approach}~(a), includes a single \gls{tfe} that processes the historical data horizon.
\begin{equation}
\label{eq:x_in_features_hist}
\mathbf{x}_{t} = \mathbf{x}_{h,t} = 
\left[
\begin{array}{r}
p^L_{t - T_{h}: t - \Delta t} \\
p^{PV}_{t - T_{h}: t - \Delta t} \\
\sin(I^{hod}_{t - T_{h}: t - \Delta t}) \\
\cos(I^{hod}_{t - T_{h}: t - \Delta t}) \\
\sin(I^{dow}_{t - T_{h}: t - \Delta t}) \\
\cos(I^{dow}_{t - T_{h}: t - \Delta t}) \\
\sin(I^{doy}_{t - T_{h}: t - \Delta t}) \\
\cos(I^{doy}_{t - T_{h}: t - \Delta t}) \\
I^{Weekend}_{t - T_{h}: t - \Delta t}
\end{array}
\right],
\end{equation}
where $\mathbf{x}_{h,t}$ contains the raw historical load $p^L$ and \gls{pv} generation $p^{PV}$, augmented with time-based embeddings encoding periodic temporal structure. The hour-of-day, day-of-week, and day-of-year indices are expressed using sine and cosine functions, while ${I^{Weekend}}$ indicates weekends. Such cyclic sine--cosine encodings are commonly used to preserve the periodic structure of temporal variables in energy time-series forecasting and have also been shown to improve learning performance in battery dispatch applications~\cite{voyant2026}.

The second variation R2D\_2TFE, shown in Figure~\ref{fig:benchmarks_proposed_approach}~(b), includes two parallel \gls{tfe}s that process historical and future information separately. The historical input $\mathbf{x}_{h,t}$ is identical to that in R2D-1TFE, while the future input $\mathbf{x}_{f,t}$ contains electricity prices over the foresight horizon and corresponding temporal embeddings:
\begin{equation}
\label{eq:x_in_features_future}
\mathbf{x}_{f,t} = 
\left[
\begin{array}{r}
{\rho}^{\text{ToU}}_{t : t + T_f} \\
\sin(I^{hod})_{t : t + T_f} \\
\cos(I^{hod})_{t : t + T_f} \\
\sin(I^{dow})_{t : t + T_f} \\
\cos(I^{dow})_{t : t + T_f} \\
\sin(I^{doy})_{t : t + T_f} \\
\cos(I^{doy})_{t : t + T_f} \\
I^{Weekend}_{t : t + T_f}
\end{array}
\right],
\end{equation}
and the combined model input is
\begin{equation}
\label{eq:x_in_combined}
\mathbf{x}_{t} = [\mathbf{x}_{h,t}, \mathbf{x}_{f,t}].
\end{equation}

\begin{table}[t]
\caption{Summary of the two R2D variants.}
\label{tab:r2d_variants}
\centering
\scriptsize
\begin{tabular}{lll}
\toprule
& 1TFE & 2TFE \\
\midrule
$\mathbf{x}_{h,t}$ & load, \gls{pv}, calendar & same \\
$\mathbf{x}_{f,t}$ & --- & \gls{tou}, calendar \\
Horizon & $T_h$ & $T_h$, $T_f$ \\
Use case & tariff unknown ahead & tariff published ahead \\
\bottomrule
\end{tabular}
\end{table}

In this work, we implement \gls{lstm}-, \gls{tcn}-, and Transformer-based \gls{tfe} variants, and recommend the \gls{lstm}-based \gls{tfe} as the default backbone due to its consistently robust performance, training stability, and computational efficiency across all tested configurations. The load and \gls{pv} generation signals exhibit relatively regular short- and medium-term temporal patterns, for which \gls{lstm}s provide sufficiently expressive representations without the overhead of more complex sequence models such as Transformers. As a result, \gls{lstm}s enable fast and reliable training while maintaining the modeling capacity required for the \gls{ems} task.

The temporal block processes the sequential input $\mathbf{x}_t$ using $n_{\mathrm{LSTM}}$ stacked \gls{lstm} layers, each with $d_{\mathrm{LSTM}}$ hidden units, to extract compact temporal representations. Depending on the information source, the encoder maps the historical input sequence $\mathbf{x}_t^{h}$ and the future input sequence $\mathbf{x}_t^{f}$ to the corresponding latent embeddings $\mathbf{z}_t^{h}$ and $\mathbf{z}_t^{f}$, respectively. The resulting latent representations are concatenated and passed to the policy module, implemented as a multi-layer perceptron (MLP) with $n_{\mathrm{MLP}}$ hidden layers, each consisting of $d_{\mathrm{MLP}}$ units. The policy network maps the fused latent features to the control action, serving as a flexible function approximator for the energy management system (\gls{ems}) decision policy.

\subsubsection{Expert Demonstration}
\label{sec:demo}
We adopt \gls{bc} as the training strategy for the proposed approach. 
The overall process consists of two main steps: demonstration data generation via \gls{milp} optimization, and supervised policy learning using the generated data.

For the training dataset, we solve an \gls{milp} problem that minimizes the total energy and battery aging costs. All battery-related degradation and thermal effects are embedded into the \gls{milp} via static cost coefficients obtained through linearization of the underlying electro-thermal and aging models. These coefficients are computed offline by evaluating the incremental \gls{soh} degradation at reference operating points and mapping it to equivalent monetary losses. This transformation enables the representation of otherwise nonlinear aging dynamics as linear cost terms within the optimization framework. The objective function for the \gls{milp} problem is:
\begin{equation}
\begin{aligned}
\label{eq:1dmilp_objective}
\min. \quad J^{1D-MILP}
= \mathbb{C}^{E} + \mathbb{C}^{B,cal} +\mathbb{C}^{B,cyc} 
\end{aligned}
\end{equation}
where
\begin{equation}
\mathbb{C}^{E}
= \sum_{t\in T}(p_t^{G,buy} \cdot k_t^{ToU} \cdot \Delta t)
- \sum_{t\in T}(p_t^{G,sell} \cdot k_t^{Sell} \cdot \Delta t),
\end{equation}
\begin{equation}
\mathbb{C}^{B,cal} =  \sum_{t\in T} (c_{cal}^{base} + c_{cal}^{T} T_t^B + c_{cal}^{soc}\frac{E_t^B}{E^N}),
\end{equation}

\begin{equation}
\mathbb{C}^{B,cyc} = \sum_{t\in T}(c_{cyc}^{ch}p_t^{B,ch}+c_{cyc}^{dch}p_t^{B,dch})\Delta t,
\end{equation}

The coefficients $c_{cal}^{T}$, $c_{cal}^{soc}$, $c_{cyc}^{ch}$, and $c_{cyc}^{dch}$ denote marginal economic degradation costs associated with temperature- and state-of-charge-dependent calendar aging, and charge- and discharge-throughput-dependent cyclic aging, respectively, and are obtained by linearizing the aging model of~\cite{schimpe2018comprehensive} implemented in the simulation environment (Appendix~\ref{sec:BESS sim}). They are scaled by the battery investment cost $\rho^B$, such that a higher $\rho^B$ increases the degradation penalty in the objective, thereby promoting more conservative battery operation.

Inverter loss is modeled using Big M method set of constraints creating a mixed-integer problem. As shown in \eqref{eq:bigM1}--\eqref{eq:bigM3} a large constant \(M^{\mathrm{inv}}\) equal to battery nominal power, $p^{N}$, and a small tolerance \(\epsilon^{\mathrm{inv}}\) link the battery power \(p^{B}_t\) with the binary variable \(b^{\mathrm{inv}}_t\)~\cite{Tanjavooru.2026}.

\begin{equation}
p^{B,ch/dch}_t \leq M^{inv} \cdot b^{inv}_{t} + \epsilon^{inv},
\label{eq:bigM1}
\end{equation}
\begin{equation}
b^{inv}_{t} = 0 \Rightarrow  p^{inv,loss,ch/dch}_t \leq \epsilon,
\label{eq:bigM2}
\end{equation}
\begin{equation}
b^{inv[m]}_{t} = 1 \Rightarrow  p^{inv,loss,ch/dch}_t: LUT,
\label{eq:bigM3}
\end{equation}

When the inverter is “off” (\(b^{inv}_{t}=0\)), this forces the battery power to (approximately) zero; when “on” (\(b^{inv}_{t}=1\)), the inverter loss \(p^{inv,ch/dch}_t\) is modeled to follow the \gls{lut} as shown in Fig.~\ref{fig:inverter} of Section~\ref{sec:BESS sim}. 

The DC-side battery charging and discharging powers are determined by accounting for inverter losses as follows:
\begin{equation}
p_t^{B,ch,dc} = p_t^{B,ch} - p_t^{inv,ch}, \quad
p_t^{B,dch,dc} = p_t^{B,dch} + p_t^{inv,dch},
\end{equation}

Based on the DC-side power flow, the thermal power generation is approximated using a linear heat-loss proxy. $\gamma^{ch}$ and $\gamma^{dch}$ represent the charging and discharging thermal loss coefficients, respectively.
\begin{equation}
p_t^{heat} =
\gamma^{ch} p_t^{B,ch,dc} +
\gamma^{dch} p_t^{B,dch,dc},
\end{equation}

Subsequently, the battery energy state is updated considering charging power, discharging power, and thermal losses. The recursive energy balance is expressed as

\begin{equation}
E_t^B
=
E_{t-1}^B
+
\Delta t
\left(
p_t^{B,ch,dc}
-
p_t^{B,dch,dc}
-
p_t^{heat}
\right), \forall t \in T,
\end{equation}
Within the \gls{milp} formulation, battery temperature dynamics are modeled using a 0D lumped thermal model, where temperature evolution is driven by internal heat generation and heat exchange with the ambient environment. The parameters $k_1$ and $k_2$ depend on the mass, specific heat, density, and volume of the battery string.
\begin{equation}
T_t^B = T_{t-1}^B + \Delta t \cdot
\left(k_1 \cdot p_t^{heat}
- k_2 \cdot (T_{t-1}^B - T_{\text{air}}) \right),
\label{eq:lumped_thermal}
\end{equation}

Binary decision variables are introduced to enforce mutually exclusive battery and grid operating modes. The variables $u_t^{B,ch}$ and $u_t^{B,dch}$ activate charging and discharging operation, respectively:
\begin{equation}
0 \leq p_t^{B,ch} \leq\,p^N u_t^{B,ch}, \forall t \in T,
\end{equation}
\begin{equation}
0 \leq p_t^{B,dch} \leq\,p^N u_t^{B,dch}, \forall t \in T,
\end{equation}

The following constraint prevents simultaneous charging and discharging:
\begin{equation}
u_t^{B,ch} + u_t^{B,dch} \leq 1, \forall t \in T,
\end{equation}

Similarly, binary variables $u_t^{G,buy}$ and $u_t^{G,sell}$ are used to enforce mutually exclusive grid import and export operation:
\begin{equation}
0 \leq p_t^{G,buy} \leq M^{grid}u_t^{G,buy}, \forall t \in T,
\end{equation}
\begin{equation}
0 \leq p_t^{G,sell} \leq M^{grid}u_t^{G,sell}, \forall t \in T,
\end{equation}
\begin{equation}
u_t^{G,buy} + u_t^{G,sell} \leq 1, \forall t \in T,
\label{eq:constrain_end}
\end{equation}

The overall AC-side power balance remains as following:
\begin{equation}
p_t^{G,buy} + p_t^{PV} + p_t^{B,dch} = p_t^{G,sell} + p_t^{L} + p_t^{B,ch}, \forall t \in T,
\end{equation}

The resulting solution represents an expert battery scheduling demonstration under perfect foresight within the linearized model. Although the \gls{milp} relies on piecewise-linear approximations and therefore does not fully capture the nonlinearities of the full simulation, it provides economically structured and aging-aware expert trajectories. These trajectories are well suited to guide the learning process, as the model accounts for key operational effects, including inverter losses, thermal losses, and aging-related constraints.

By solving the \gls{lp} problem for the input time series 
$p^L_{t_{\text{start}}:t_{\text{end}}}$, 
$p^{PV}_{t_{\text{start}}:t_{\text{end}}}$, and 
${\rho}^{\text{ToU}}_{t_{\text{start}}:t_{\text{end}}}$, we obtain the optimal battery charging and discharging schedule:
\begin{equation}
p^{B,\text{1D-MILP}}_{t_{\text{start}}:t_{\text{end}}} 
= p^{B,\text{ch}}_{t_{\text{start}}:t_{\text{end}}} 
- p^{B,\text{dch}}_{t_{\text{start}}:t_{\text{end}}}.
\end{equation}

At each time step $t$, we construct one data point whose input features $x_t$ correspond to Eqs.~\eqref{eq:x_in_features_hist} and 
\eqref{eq:x_in_features_future} for two proposed architectures, respectively, and whose label is the \gls{milp}-based battery power setpoint $\hat{\mathbf{y}}_{t} =p_{t:t+T_c}^{B,\text{1D-MILP}}$.

\subsubsection{Behavior Cloning} 
The \gls{r2d} network is trained to imitate the LP-derived reference actions by minimizing the \gls{mse} between the predicted action $\pi_\theta(x_t)$ and the label $\hat{\mathbf{y}}_{t}$ provided by MILP Expert setpoint: 

\[
    \mathcal{L}_{\text{BC}}(\theta) = \frac{1}{T} \sum_{t=1}^{T}
    \bigl\| \pi_\theta(x_t) - \hat{\mathbf{y}}_{t} \bigr\|^2.
\]

Training is performed for $n_{\text{BC}}$ epochs.  
After each epoch, the validation loss is computed on the validation dataset, and the model achieving the lowest validation loss is selected as the final policy.

\subsubsection{Training and Deployment Workflow}
\label{sec:workflow}

R2D operates in two phases that share the same input encoding but differ in how decisions are produced. In the offline phase, the aging aware MILP expert solves each historical window to optimality over the full information horizon and returns the battery setpoints, which serve as supervision labels; the policy $\pi_\theta$ is then trained by behavior cloning to reproduce these labels by minimizing $\mathcal{L}_{\mathrm{BC}}$. In the online phase, the trained policy runs forward only, with no optimization solved and no explicit forecast generated at runtime. At each control step the controller observes a historical window of length $T_h$ for load, PV, and battery states together with the day-ahead price and calendar indices over the foresight horizon $T_f$, following the horizon definition in Fig.~\ref{fig:horizon}; each stream is processed without further normalization by its modular \gls{tfe} into the latent embeddings $\mathbf{z}^{h}_t$ and $\mathbf{z}^{f}_t$, which the decision network maps to the AC side setpoint $p^{B,AC}_{t:t+T_c}$ applied to the BESS environment to advance the state, as shown in Fig.~\ref{fig:usecase} with the encoder structure in Fig.~\ref{fig:benchmarks_proposed_approach}. Because the expert schedule already embeds aging, inverter, and thermal knowledge, this knowledge is transferred to R2D entirely through the cloned state to action mapping, without the online policy ever solving or forecasting explicitly.

\subsection{Experiments}
\label{sec:experiments}
In this section, we discuss the experiment setup, including the simulation environment, data selection, and parameters for all benchmarking models. 

\subsubsection{Simulation Environment}

All experiments are implemented in the public Git repository \texttt{R2D}~\cite{R2D2026}. 
The energy system, including detailed battery simulation, is implemented as a Gym-based environment that supports interaction with the proposed \gls{r2d} controller as well as RL- and \gls{mpc}-based baselines. 

The simulation environment provides historical observations to the controller at each iteration according to the selected time-horizon configuration. After the control decision is computed, the resulting action is applied to the environment to advance the system state. If the control horizon exceeds a single time step, the computed control sequence is dispatched sequentially within the simulation.

\subsubsection{Data}
\label{sec:data}

In this work, consumption and \gls{pv} generation data are obtained from the EMSx dataset~\cite{le2023emsx}, which initially contains 70 sites with heterogeneous battery sizing and partly discontinuous records. From these, we selected sites with continuous time series data spanning at least two years and two months. Five sites (IDs 10, 11, 12, 13, and 70) met this criterion. The load and \gls{pv} generation data are normalized to remove scale effects by rescaling according to a nominal battery size fixed at 100~kWh, ensuring comparability across sites, following the procedure of~\cite{yin2025boosting}. The dataset provides one measured \gls{pv} series rescaled to each site, so the five profiles differ in load shape and in \gls{pv} and battery sizing rather than in \gls{pv} shape; after normalization the mean load ranges from 8.7 to 38.3~kW, the peak load from 62.7 to 66.6~kW, the annual \gls{pv}-to-load energy ratio from 0.35 to 1.83, and the \gls{tou} price averages 0.166~€/kWh with a standard deviation of 0.023~€/kWh.

\gls{tou} prices are derived from the \gls{dam} \cite{EnergyCharts2026} and represent dynamic electricity prices converted into ToU tariffs over the same period, reflecting typical industrial energy price levels. The fixed feed-in tariff $k^{\text{Sell}} = 0.086$~€/kWh is set in accordance with the German Renewable Energy Sources Act (EEG), which guaranteed remuneration rates of approximately 8--9~ct/kWh for small-to-medium \gls{pv} installations during the 2015--2017 study period~\cite{yin2025boosting}.

\begin{figure}[htbp]
\centering
\includegraphics[width=1\columnwidth]{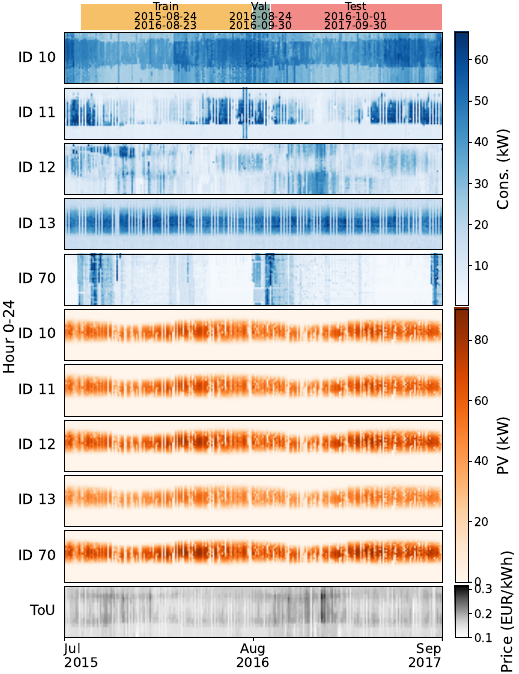}
\caption{Heatmap visualization of the dataset split and input profiles for all five industrial sites \cite{le2023emsx}. Each row represents hourly data over the full two-year period (Aug~2015--Sep~2017), with training, validation, and test splits indicated at the top. Blue panels show electricity consumption (kW), orange panels show \gls{pv} generation (kW), and the bottom row shows the \gls{tou} electricity price (€/kWh).}
\label{fig:data}
\end{figure}

The combined dataset spans two years and two months, divided into one year for training, one month for validation, and one year for testing, as visualized in Fig.~\ref{fig:data}. This split reflects a realistic industrial deployment scenario: assuming a decision point of 30~September~2016, all models are trained and selected exclusively on historically available data (August~2015--September~2016), and evaluated on the subsequent out-of-sample year (October~2016--September~2017), which contains 35{,}040 steps at 15~min resolution with no missing values. Both training and test sets span at least one full calendar year, ensuring that all seasonal load, \gls{pv}, and price patterns are represented in each split.

\subsubsection{Benchmark Study}
\label{sec:benchmarks}

This benchmark study compares the proposed R2D against established controllers in a method-wise manner: each method, including R2D, is represented by a single fixed configuration, and all controllers are evaluated under identical simulation conditions, data splits, and metrics.

We evaluate the proposed \gls{r2d} variants against six benchmark controllers, which are summarized in Table~\ref{tab:benchmarks}. They span clairvoyant, optimization-based, learning-based, and hybrid families. Two \emph{clairvoyant} baselines use ground-truth \gls{pv}, load, and price data to solve the \gls{milp} defined in Equations~\eqref{eq:1dmilp_objective}--\eqref{eq:constrain_end}: Global\_CF solves once over the full dataset horizon (theoretical upper bound), while MPC\_CF re-solves at every step within a limited receding horizon (upper bound under perfect short-term foresight). Four \emph{realistic} controllers operate without ground-truth future information. Three \gls{mpc} variants follow a predict-then-optimize loop using different forecasting strategies: MPC\_PD uses persistent deterministic forecasts (rolling average over the previous $n_h = T_h/24\,\text{h}$ days); MPC\_ML is a hybrid \gls{mpc} controller that reuses the \gls{r2d} \gls{tfe} architecture as a multi-step forecaster, where the policy head is replaced by a linear projection head trained to minimize \gls{mse} over the foresight horizon $T_f$; and MPC\_PQ employs nine persistent quantile forecasts (10\%--90\%) in a scenario-based stochastic \gls{lp} with a here-and-now battery decision structure. Finally, RL\_STD is a \gls{ppo} agent with an \gls{mlp} policy that learns implicitly from historical interactions without explicit forecasts. Its reward is defined as the cost saving minus the battery aging cost,
\begin{equation}
    r_t = c^E_t - \bar{c}^E_t -c^{B}_t ,
\end{equation}
where $\bar{c}^E_t$ denotes the energy cost of the no-battery baseline, and $c^{B}_t$ denotes the aging cost obtained by converting the incremental \gls{soh} loss into a monetary penalty using the battery investment value $\rho^B$.

For the benchmark study, both R2D variants use a fixed default configuration: an LSTM temporal backbone of size S (detailed in Section~\ref{sec:sensitivity_study}), the 1D-MILP teacher introduced in Section~\ref{sec:demo}, a 72~h historical horizon (with an additional 24~h foresight horizon for 2TFE only), and a single step control horizon of 15~min. This default is held constant across all five sites.
\begin{table*}[ht]
\centering
\caption{Overview of benchmark controllers.}
\label{tab:benchmarks}
\begin{tabular}{lccl}
\toprule
\textbf{Controller} & \textbf{Forecast Type} & \textbf{Decision-Making} & \textbf{Role} \\
\midrule
Global\_CF  & Perfect (full dataset)                              & \gls{milp}          & Global upper bound    \\
MPC\_CF     & Perfect (receding)                                  & \gls{milp}          & \gls{mpc} upper bound \\
MPC\_PD     & Persistent deterministic                            & \gls{milp}          & \gls{mpc} baseline    \\
MPC\_ML     & Data-driven (\gls{r2d}-\gls{tfe}s as backbone) & \gls{milp}          & Hybrid baseline       \\
MPC\_PQ     & Persistent quantiles                                & Scenario \gls{lp}   & Stochastic baseline   \\
RL\_STD     &  No forecast                              & \gls{ppo}/\gls{mlp} & \gls{rl} baseline     \\
\bottomrule
\end{tabular}
\end{table*}

\subsubsection{Sensitivity Study}
\label{sec:sensitivity_study}

This sensitivity study analyzes how the proposed R2D responds to its main design choices. It isolates the effect of one factor at a time while all other factors are held at the default configuration defined in Section~\ref{sec:benchmarks}, in order to characterize the robustness and behavior of R2D. Three factors are examined: neural network architecture, expert quality, and time horizon configuration.

First, we conducted an architecture ablation study for the proposed \gls{r2d} policy. The tested architecture dimensions and detailed model presets are summarized in Tables~\ref{tab:r2d_ablation_space} and \ref{tab:r2d_ablation_presets}. This experiment was conducted to provide further insight into how neural network architecture affects \gls{r2d} controllers.

\begin{table}[!t]
\caption{Architecture variations considered in the R2D ablation study.}
\label{tab:r2d_ablation_space}
\centering
\begin{tabular}{lc}
\toprule
\textbf{Component} & \textbf{Options} \\
\midrule
R2D variant & \{1TFE / 2TFE\} \\
Temporal backbone & \{LSTM / TCN / Transformer\} \\
Model size & \{S / M / L / XL\} \\
\bottomrule
\end{tabular}
\end{table}

\begin{table*}[htbp]
\caption{Detailed architecture presets used in the R2D ablation study. In 1TFE, only the historical encoder is active. In 2TFE, both the historical and future encoders are active, with the same backbone family and the preset dimensions shown below. TCN channel lists are abbreviated as $n\times c$ for $n$ residual blocks of $c$ channels each. The last column gives the number of trainable parameters of the 1TFE and the 2TFE variant}
\label{tab:r2d_ablation_presets}
\centering
\scriptsize
\setlength{\tabcolsep}{4pt}
\begin{tabular}{lllllr}
\toprule
TFE Type & Size & Historical TFE ($ts_h$) & Future TFE ($ts_f$) & Decision network $(\pi)$ & \#Param. \\
\midrule
LSTM & S  & hidden $=8$, layers $=1$, $z^{h}=8$   & hidden $=8$, layers $=1$, $z^{f}=8$   & $[8]$ & 20\,355 / 21\,955 \\
LSTM & M  & hidden $=16$, layers $=2$, $z^{h}=16$ & hidden $=8$, layers $=1$, $z^{f}=8$   & $[20]$ & 27\,795 / 29\,395 \\
LSTM & L  & hidden $=28$, layers $=1$, $z^{h}=16$ & hidden $=16$, layers $=1$, $z^{f}=16$ & $[16,16]$ & 28\,227 / 31\,939 \\
LSTM & XL & hidden $=40$, layers $=1$, $z^{h}=24$ & hidden $=16$, layers $=2$, $z^{f}=16$ & $[32,20,10]$ & 38\,875 / 44\,763 \\
\midrule
TCN & S  & ch. $=6\times4$, $k=5$, $z^{h}=8$   & ch. $=4\times4$, $k=5$, $z^{f}=8$   & $[8]$ & 20\,935 / 22\,787 \\
TCN & M  & ch. $=6\times8$, $k=5$, $z^{h}=16$  & ch. $=4\times8$, $k=5$, $z^{f}=8$   & $[20]$ & 28\,091 / 31\,739 \\
TCN & L  & ch. $=8\times8$, $k=3$, $z^{h}=16$  & ch. $=5\times8$, $k=3$, $z^{f}=16$  & $[16,16]$ & 26\,843 / 31\,035 \\
TCN & XL & ch. $=9\times10$, $k=5$, $z^{h}=20$ & ch. $=5\times10$, $k=5$, $z^{f}=16$ & $[32,20,10]$ & 38\,669 / 45\,983 \\
\midrule
Transformer & S  & $d=12$, heads $=2$, layers $=1$, ff $=32$, $z^{h}=12$ & $d=8$, heads $=2$, layers $=1$, ff $=16$, $z^{f}=8$  & $[8]$ & 21\,863 / 23\,559 \\
Transformer & M  & $d=16$, heads $=2$, layers $=2$, ff $=64$, $z^{h}=16$ & $d=12$, heads $=2$, layers $=1$, ff $=32$, $z^{f}=12$ & $[20]$ & 30\,611 / 33\,739 \\
Transformer & L  & $d=20$, heads $=2$, layers $=2$, ff $=64$, $z^{h}=20$ & $d=16$, heads $=2$, layers $=1$, ff $=48$, $z^{f}=16$ & $[16,16]$ & 32\,915 / 37\,859 \\
Transformer & XL & $d=20$, heads $=2$, layers $=3$, ff $=64$, $z^{h}=20$ & $d=16$, heads $=2$, layers $=2$, ff $=64$, $z^{f}=16$ & $[32,20,10]$ & 42\,631 / 51\,383 \\
\bottomrule
\end{tabular}
\end{table*}

Second, we conducted an expert-sensitivity experiment for behavior cloning. In addition to the aging-aware 1D-MILP expert presented in Section~\ref{sec:demo}, we implemented two additional optimization-based teachers with lower modeling fidelity: an LP expert that considers only electricity cost, and an aging-aware MILP expert with simplified battery degradation modeling. The main differences among the tested expert formulations are summarized in Table~\ref{tab:expert_models}, while the corresponding formulations of the LP and MILP experts are provided in the Appendix. For the aging-aware expert policies, the battery replacement value $\rho^B$ was further varied by setting matched energy- and power-based investment cost coefficients to 0, 100, 200, 300, and 400~€/kWh, respectively. No such parameter is required for the purely economic LP expert. This experiment was conducted to examine how expert formulation and multi-objective sensitivity, through the battery replacement value and the weighting of aging cost in the objective function, affect economic performance, \gls{soh} preservation, and efficiency.

\begin{table}[t]
\caption{Modeling aspects of the expert formulations considered in the expert-sensitivity study.}
\label{tab:expert_models}
\centering
\begin{tabular}{lccc}
\toprule
Model component & LP & MILP & 1D-MILP \\
\midrule
Energy-cost optimization & \checkmark & \checkmark & \checkmark \\
Calendar aging &  & \checkmark & \checkmark \\
Cycling aging &  & \checkmark & \checkmark \\
Inverter losses &  &  & \checkmark \\
Thermal dynamics &  &  & \checkmark \\
\bottomrule
\end{tabular}
\end{table}

Third, we conducted a horizon-sensitivity study for the \gls{r2d} policy. This experiment examined how the learned controller responds to changes in the historical observation horizon, foresight horizon, and control horizon. The tested settings are summarized in Table~\ref{tab:r2d_horizon_sensitivity}. In each test, the factor under study is varied while the remaining horizons are held at the default.

\begin{table}[t]
\caption{Varied horizon settings in the R2D horizon-sensitivity study. For each test, the non-varied horizons were fixed.}
\label{tab:r2d_horizon_sensitivity}
\centering
\begin{tabular}{lc}
\toprule
Experiment & Tested values \\
\midrule
Historical horizon $H$ & $\{24,\,72,\,120,\,168\}$ \\
Foresight horizon $F$ & $\{1,\,4,\,8,\,12,\,16,\,20,\,24\}$ \\
Control horizon $C$ & $\{0.25,\,1,\,4,\,8,\,12,\,16,\,20,\,24\}$ \\
\bottomrule
\end{tabular}
\end{table}

\subsubsection{Out-of-Distribution Test}
\label{sec:ood_setup}
Beyond the benchmark and sensitivity studies, we assess the generalization of R2D to unseen data through an \gls{ood} test. The trained policies, using the fixed default configuration, are evaluated without any retraining under two protocols. The cross-site protocol evaluates each policy on the complete test profiles of all five sites, measuring how well a policy learned on one load, PV, and price profile transfers to profiles it never observed during training, reflecting a realistic deployment in which a controller is reused across sites without retraining. The single-factor protocol instead substitutes one exogenous input at a time while the remaining inputs stay those of the target profile: the load shape of site~13, the measured 2020 FMI Helsinki PV profile~\cite{Karhu2026}, or the 2021 German day-ahead price series~\cite{EnergyCharts2026}. Each substituted series is calendar aligned, normalized to the peak, respectively, the tariff range, of the target profile, and additionally scaled by 0.7, 1.0 and 1.3, so that a loss of performance can be attributed to a single input. The economic performance under this test is reported in Section~\ref{sec:results_ood}.

\subsubsection{Parameters}
\label{sec:para}

The key simulation and training parameters are summarized in Table~\ref{tab:para}, with full \gls{bess} parameters provided in Appendix~\ref{sec:values} (Table~\ref{tab:bess_params}). The architectural parameters of the \gls{r2d} model, including \gls{tfe} and decision layer configurations across model sizes, are summarized in Table~\ref{tab:r2d_ablation_presets}. The \gls{r2d} policy network is trained with a batch size of 32, a minimum of 20 and a maximum of 100 epochs, and an early stopping criterion of 10 consecutive epochs without improvement. All stochastic models are evaluated across 10 independent seeds to account for training variability. 

\begin{table}[htbp]
    \centering
    \caption{Parameter configuration for battery simulation and the associated optimization model used for demonstration generation in imitation learning.}
    \begin{tabular}{llc}
    \toprule
    \textbf{Parameter} & \textbf{Symbol} & \textbf{Value} \\
    \midrule
    Battery capacity (kWh)                    & $E^N$                  & 100 \\
    Battery maximum power (kW)                & $p^N$                  & 100 \\
    Initial state of charge                   & $soc_{\text{start}}$   & 0.1 \\
    Minimum state of charge                   & $soc_{\min}$           & 0.1 \\
    Maximum state of charge                   & $soc_{\max}$           & 0.9 \\
    Initial state of health                   & $soh_{\text{start}}$   & 1.0 \\
    Time step (h)                             & $\Delta t$             & 0.25 \\
    \bottomrule
    \end{tabular}
    \label{tab:para}
\end{table}

\subsubsection{Metrics}
\label{sec:metrics}

All results are reported on the test dataset. We evaluate each controller using three key performance indicators: economic performance, battery degradation, and efficiency.

Economic performance is measured by the saving of model $M$,
\begin{equation}
S_M = \mathbb{C}_{\mathrm{Net}} - \mathbb{C}_M,
\end{equation}
where $\mathbb{C}_{\mathrm{Net}}$ is the total electricity cost without battery operation and $\mathbb{C}_M$ is the total electricity cost under model $M$. To assess how closely a controller approaches strong reference solutions, we additionally normalize the achieved saving by two upper-bound benchmarks,
\begin{equation}
\bar{S}^{B}_M = \frac{S_M}{S_B}, \qquad B \in \{Global\_CF,\; MPC\_CF\},
\end{equation}
where $Global\_CF$ denotes the full-horizon clairvoyant benchmark and $MPC\_CF$ denotes the clairvoyant receding-horizon MPC benchmark.

Battery degradation is measured by the state-of-health loss over the evaluation horizon,
\begin{equation}
\Delta soh_M = soh_{t_1} - soh_{t_{\mathrm{end}}},
\end{equation}
where $soh_{t_1}$ and $soh_{t_{\mathrm{end}}}$ are the initial and final battery state of health.

\section{Results}
\label{sec:results}

In this section, we present the results of the benchmark study, the sensitivity study, and the \gls{ood} test. The benchmark study reports the annual performance of R2D against the six benchmarks, followed by a representative weekly operation view. The sensitivity study then analyzes how R2D responds to neural network architecture, expert quality, and time horizon. Finally, the \gls{ood} test evaluates cross-site generalization.

\subsection{Benchmarking Study Results}
\label{sec:results_kpi}

\subsubsection{Annual Analysis}
The normalized economic savings $S$ and annual \gls{soh} loss $\Delta SOH$ of all controllers across five industrial sites are summarized in Fig.~\ref{fig:2kpi}. Detailed numeric results are reported in Appendix~\ref{sec:full_results}.

\begin{figure*}[t]
    \centering
    \includegraphics[width=1\textwidth]{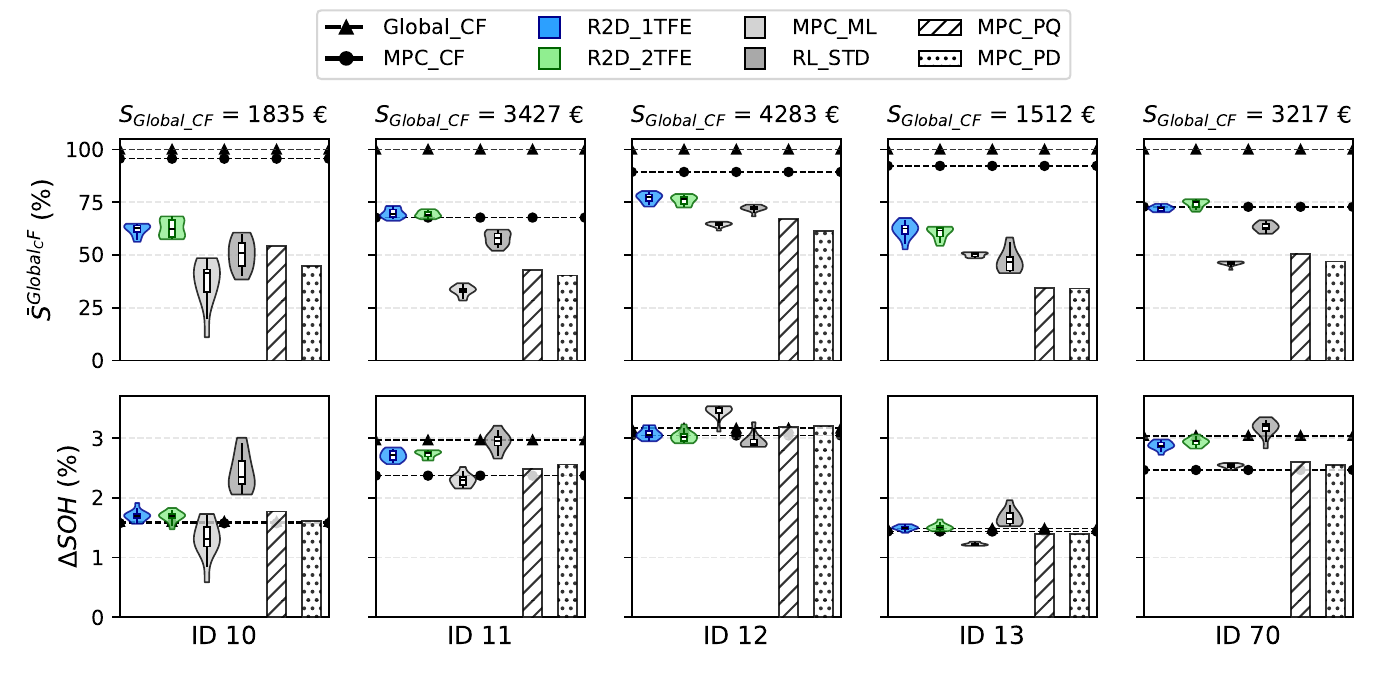}
    \caption{Normalized savings $\bar{S}^{B}$, annual \gls{soh} loss $\Delta$SOH of the proposed \gls{r2d} controllers (R2D\_1TFE in blue, R2D\_2TFE in green) compared with clairvoyant benchmarks (Global\_CF: dashed line with triangles; MPC\_CF: dashed line with dots) and realistic benchmarks (MPC\_ML, RL\_STD, MPC\_PQ, MPC\_PD) across five industrial sites. The absolute savings of Global\_CF are reported above each column. Controllers involving stochastic training are evaluated over ten independent runs and shown as violin plots; deterministic controllers are shown as bar plots.}
    \label{fig:2kpi}
\end{figure*}

The absolute savings of Global\_CF range from 1512~€ for ID~13 to 4283~€ for ID~12, indicating substantial site-specific variability. MPC\_CF achieves between 68\% and 96\% of the global optimum depending on the site. The highest MPC\_CF performance is observed for ID~12, where it reaches 89\% of the global optimum, while lower relative performance is observed for profiles such as IDs~11 and~70.

Among all realistic controllers, R2D\_1TFE and 2TFE achieve the highest normalized savings across the five sites, attaining 62\%--77\% of the global optimum. R2D\_1TFE achieves the best performance for IDs~11, 12, and~13, whereas R2D\_2TFE performs best for IDs~10 and~70. The violin plots in Fig.~\ref{fig:2kpi} show narrow distributions for both R2D variants, with a standard deviation of at most 4.4 percentage points (pp) across all sites.

The MPC baselines show larger variation across sites. MPC\_PD provides the lowest performance among the realistic controllers, reaching 34\%--62\% of the global optimum. MPC\_PQ achieves 34\%--67\% of the global optimum and is the strongest MPC-based baseline among the tested variants. Across all sites, R2D exceeds MPC\_PQ by 2\%--30\%. MPC\_ML improves substantially over MPC\_PD in several cases, including ID~13, where its savings increase from near-zero to 50\%. However, its performance varies across sites, ranging from 33\% for ID~11 to 64\% for ID~12.

RL\_STD achieves at least 47\% of the global optimum across all sites. Compared with R2D, however, its violin distributions are wider, particularly for ID~10 ($\pm$7.4~pp) and ID~13 ($\pm$5.5~pp). In terms of economic savings, RL\_STD remains below R2D across most sites. RL\_STD also results in higher annual \gls{soh} loss than R2D at most sites, with differences of 0.2--0.7~pp. For example, the annual \gls{soh} loss is 2.43\% for RL\_STD compared with 1.70\% for R2D at ID~10, and 2.94\% compared with 2.71\% at ID~11. The exception is ID~12, where RL\_STD shows slightly lower degradation than R2D, with 2.96\% compared with 3.08\%.

Regarding battery degradation, the optimization-based controllers, including the clairvoyant benchmarks, show annual \gls{soh} losses between 1.5\% and 3.2\%. R2D achieves comparable degradation levels across the tested sites. RL\_STD shows elevated degradation at most sites, with ID~12 being the main exception.

\subsubsection{Weekly Operation View}
\label{sec:results_operational}

\begin{figure}[t]
    \centering
    \includegraphics[width=0.5\textwidth]{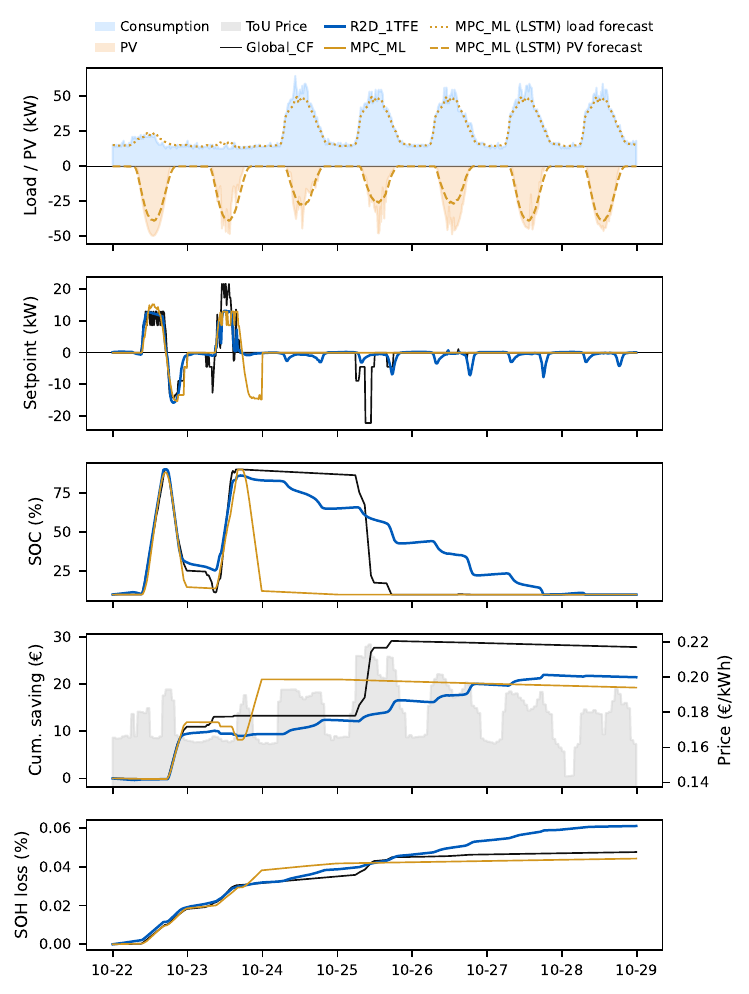}
    \caption{One-week operational profile for site ID~13 (Oct~22--29) comparing Global\_CF, R2D\_1TFE, and MPC\_ML. From top to bottom: load and \gls{pv} generation with MPC\_ML day-ahead forecasts (dotted/dashed); battery setpoint; \gls{soc}; cumulative savings overlaid with \gls{tou} price; and cumulative \gls{soh} loss. The load profile is regular and repetitive with clear daily \gls{pv} peaks, representing a favorable condition for forecast-based control.}
    \label{fig:op_profile}
\end{figure}

To provide insight into the operational behavior of the controllers, Fig.~\ref{fig:op_profile} presents a representative one-week profile for site ID~13 from Oct~22 to Oct~29. The comparison includes Global\_CF, R2D\_1TFE, and MPC\_ML.

The selected profile is characterized by a regular and repetitive load pattern with pronounced daily \gls{pv} peaks. Global\_CF charges the battery during the first two days and defers the main discharge to day~4, when the electricity price reaches a peak. This results in approximately 29~€ of cumulative savings over the selected week. R2D\_1TFE follows a more gradual discharge pattern during the high-price period and accumulates approximately 21~€ of cumulative savings. MPC\_ML discharges the battery shortly after the first two days and then remains largely idle for the rest of the selected period, because its rolling horizon forces the stored energy to be discharged before the horizon ends, whereas R2D reproduces an averaged form of the clairvoyant schedule and retains energy for the price peak on 26~October; the same limitation of our rolling-horizon \gls{mpc} implementation explains why R2D exceeds \gls{mpc}\_CF at IDs~11 and~70 in Fig.~\ref{fig:2kpi}.

The setpoint error of R2D\_1TFE relative to the Global\_CF reference is 1.32 in terms of \gls{mae} and 9.82 in terms of \gls{mse}. For MPC\_ML, the 24-step-ahead load forecast results in a \gls{mae} of 2.09 and a \gls{mse} of 9.18, while the \gls{pv} forecast results in a \gls{mae} of 2.98 and a \gls{mse} of 29.29. The resulting net-load forecast error is higher than the individual forecast errors, with a \gls{mae} of 3.67 and a \gls{mse} of 33.53.

\subsection{Sensitivity Study Results}
\label{sec:sensitivity_results}
\subsubsection{Architecture}
\label{sec:results_ablation}

\begin{figure*}[B]
    \centering
    \includegraphics[width=1\textwidth]{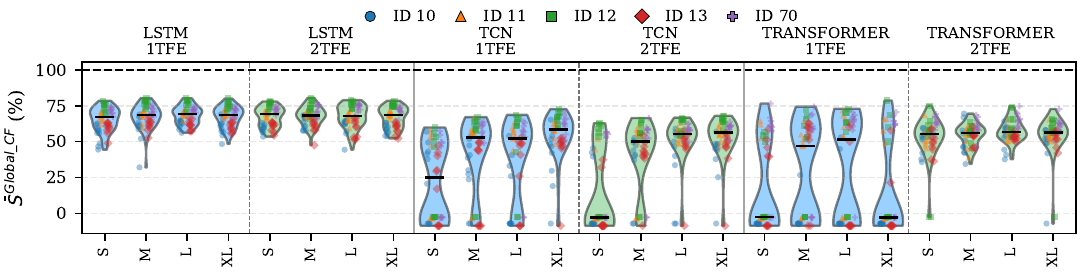}
    \caption{Normalized savings $\bar{S}^{Global\_CF}$ of all \gls{r2d} architecture combinations across five industrial sites, grouped by temporal backbone, \gls{tfe} stream configuration, and model size. Each violin aggregates results across sites and seeds; the horizontal bar denotes the median. The dashed line indicates the Global\_CF upper bound.}
    \label{fig:ablation}
\end{figure*}

Fig.~\ref{fig:ablation} presents the normalized savings of all \gls{r2d} architecture combinations across the five industrial sites. The comparison covers three temporal backbones, namely LSTM, TCN, and Transformer, two \gls{tfe} stream configurations, namely 1TFE and 2TFE, and four model sizes, namely S, M, L, and XL.

Across the tested configurations, LSTM achieves the most stable performance. Its violin distributions are tightly concentrated across model sizes, \gls{tfe} configurations, and sites, with median savings close to the best-performing configurations. The performance of LSTM is also relatively insensitive to model size.

TCN and Transformer show larger variability than LSTM. For TCN, performance improves with increasing model size, while smaller configurations exhibit lower savings and wider distributions. Transformer performance varies more strongly with the \gls{tfe} stream configuration, and the relative performance of 1TFE and 2TFE differs across sites. Overall, the ablation results show that LSTM provides the most robust performance among the tested temporal backbones.

The computational footprint of R2D is also governed by the architecture.
Detailed runtime and memory measurements for \gls{r2d}, \gls{mpc}, and \gls{rl} controllers are reported in Appendix~\ref{sec:app_runtime}. The lightest \gls{r2d} configuration, namely 1TFE with \gls{lstm}-S, requires a mean step time of 0.41~ms and an initialization memory of 127~MB.
\subsubsection{Expert}
\label{sec:results_expert}

\begin{figure*}[t]
    \centering
    \includegraphics[width=1\textwidth]{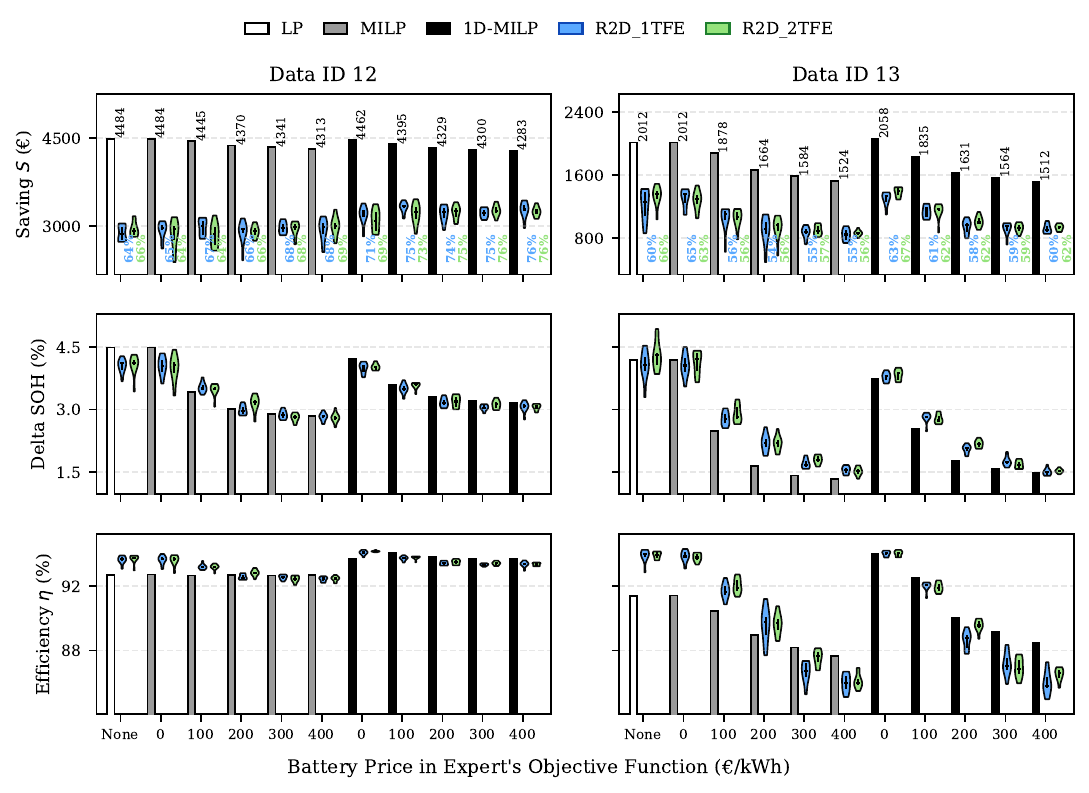}
    \caption{Detailed KPI comparison for representative Data IDs 12 and 13 under different battery-price coefficients in the expert objective function. The left column shows ID 12 and the right column shows ID 13. The three rows report, from top to bottom, saving $S$ in €, battery degradation $\Delta \mathrm{SOH}$ in percent, and efficiency $\eta$ in percent. Along the horizontal axis, each group corresponds to one battery-price coefficient. Within each group, white, gray, and black bars denote the expert optimization baselines LP, MILP, and 1D-MILP, respectively, while the blue and green violin plots denote the distributions of the learned controllers R2D\_1TFE and R2D\_2TFE. The percentages annotated in the top row indicate the saving of each R2D variant normalized by the corresponding expert baseline.}
    \label{fig:expert1213}
\end{figure*}

The sensitivity study on the expert models and the coefficients of the multi-objective function is shown in Fig.~\ref{fig:expert1213}. Three expert models are implemented, as listed in Table~\ref{tab:expert_models}. For the aging-aware MILP model, the aging weight is calculated using battery costs of 0, 100, ..., 400~€/kWh. Fig.~\ref{fig:expert1213} reports two representative cases: ID~12, which behaves similarly to ID~11 and ID~70, and ID~13, which behaves similarly to ID~10.

When the battery cost is zero, both MILP models achieve the same savings because the optimization problem becomes a single-objective cost-saving maximization. As the battery cost increases, both savings and \gls{soh} loss decrease. For ID~13, considering aging cost reduces the savings by more than 500~€, corresponding to about 25\%, while reducing \gls{soh} loss by nearly 3\%. For ID~12, the savings decrease from 4462~€ to 4283~€, corresponding to about 4\%, while \gls{soh} loss is reduced from 4.4\% to 3\%.

The R2D performance changes with the selected expert model and battery-price coefficient. When the expert with the most detailed battery representation is used as the teacher, the R2D model shows lower performance variance. Higher aging costs shift the R2D violin-plot distributions toward lower \gls{soh} loss. For ID~12 with the 1D-MILP expert, increasing the battery cost increases the normalized R2D savings from 69\% to 76\%, while the expert's own profitability decreases and the \gls{soh} loss is reduced.

\begin{figure}[t]
    \centering
    \includegraphics[width=0.48\textwidth]{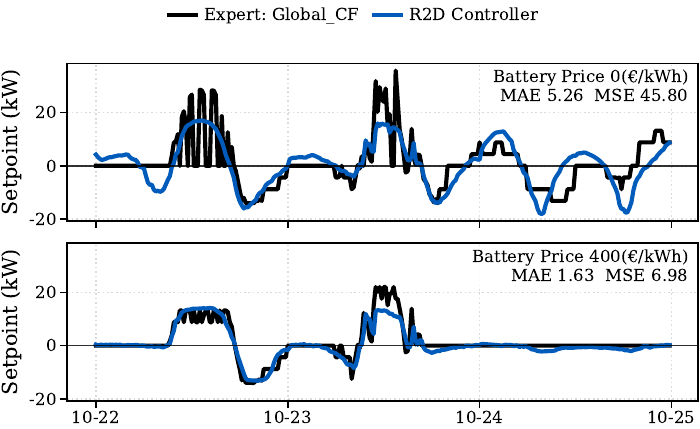}
    \caption{Comparison of battery power setpoints for Data ID 13 over a representative horizon. The black curve shows the expert Global\_CF solution and the blue curve shows the R2D controller output. The upper subplot uses battery price 0~€/kWh and the lower subplot uses 400~€/kWh in the expert objective. Reported MAE and MSE values quantify the mismatch between the learned controller and the expert reference, with the 400~€/kWh case showing substantially improved agreement.}
    \label{fig:expert_0_400}
\end{figure}

Fig.~\ref{fig:expert_0_400} compares the expert and R2D battery power setpoints for two representative battery-price coefficients. With a battery price of 0~€/kWh, the expert produces a more aggressive and zigzag charging pattern. With a battery price of 400~€/kWh, the expert produces a smoother setpoint trajectory with fewer actions. The smoother expert trajectory reduces the R2D fitting error from a \gls{mae} of 5.26 and a \gls{mse} of 45.8 to a \gls{mae} of 1.63 and a \gls{mse} of 6.89 over the four-day horizon.

\subsubsection{Horizon}
\label{sec:results_horizon}

The sensitivity of \gls{r2d} performance to horizon configuration is reported in Appendix~\ref{sec:app_Horizon}. Foresight and historical horizons have negligible impact on economic savings, which remain stable within $\pm$2\% across all tested values and sites. In contrast, the control horizon shows a clear effect on economic performance. Shorter control horizons lead to higher savings, while performance decreases as the control horizon extends beyond 15~min. The largest observed reduction is up to 27~pp at site~10 under a 24~h control horizon.

\subsection{Out-of-Distribution Generalization}
\label{sec:results_ood}

\begin{figure}[t]
    \centering
    \includegraphics[width=0.48\textwidth]{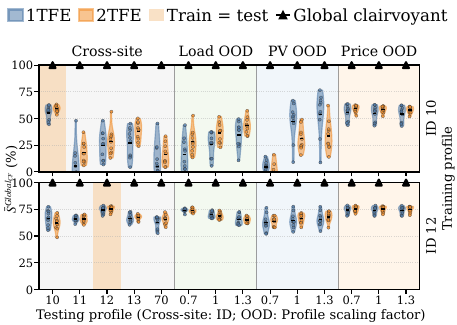}
    \caption{Robustness of the 1TFE and 2TFE controllers under cross-site and single-factor \gls{ood} (OOD) conditions. The rows correspond to models trained on profiles ID~10 and ID~12, selected as representatives of the low- and high-profitability profile groups. Cross-site evaluation uses the complete test profiles from sites 10, 11, 12, 13, and 70, with the matched train--test profile highlighted. Load OOD replaces only consumption with the site-13 load shape, PV OOD only \gls{pv} generation with the calendar-aligned 2020 FMI Helsinki profile~\cite{Karhu2026}, and price OOD only the ToU tariff with calendar-aligned 2021 day-ahead market prices~\cite{EnergyCharts2026} mapped to the original tariff range; each substituted series is normalized to the corresponding peak or tariff range of the target profile and scaled by factors of 0.7, 1.0, and 1.3. Savings are reported as the normalized saving $\bar{S}^{Global\_CF}$ defined in Section~\ref{sec:metrics}, with the clairvoyant reference recomputed for each test profile and therefore marked at 100\% for every condition. Violin distributions summarize ten random seeds, dots indicate individual seeds, and black horizontal bars denote means.}
    \label{fig:OOD}
\end{figure}

To evaluate the generalization ability of the \gls{r2d} framework, the policies trained on ID~10 and ID~12 are evaluated under the cross-site and single-factor protocols and are shown as representatives of the tested site models to limit the figure size. The economic performance is summarized in Fig.~\ref{fig:OOD}.

The two training profiles behave very differently. The policy trained on ID~12 remains between 62\% and 75\% of the profile-specific clairvoyant optimum across all nine single-factor conditions, with a seed standard deviation of at most 7.6~pp, close to its in-distribution level of 77\%, and holds this level across all five cross-site profiles. The policy trained on ID~10 spans $-8$\% to 59\% over the same conditions, with a seed spread of up to 20.3~pp, and is unstable across sites.

The three exogenous inputs contribute unequally. The tariff is the most benign: scaling the substituted day-ahead price changes the normalized saving by less than 1.5~pp, from 74\% to 75\% for ID~12 and 54\% to 59\% for ID~10. Load substitution is tolerated by the ID~12 policy at 65--74\% but reduces the ID~10 policy to 16--43\%. \gls{pv} substitution is the most demanding: ID~12 retains 62--68\%, whereas ID~10 falls to 4.5\% for 1TFE and $-7.6$\% for 2TFE at the reduced \gls{pv} level.

When compared with the input profile characteristics in Fig.~\ref{fig:data} and the benchmark performance in Fig.~\ref{fig:2kpi}, ID~12 belongs to the high-profitability sites with frequent and diverse battery activity, and ID~10 to the low-profitability group. The \gls{ood} results indicate that models trained on high-profitability profiles transfer more effectively, and that the residual sensitivity concentrates in the \gls{pv} input.

\subsection{Latent Representation}
\label{sec:results_latent}

\begin{figure}[t]
    \centering
    \includegraphics[width=0.5\textwidth]{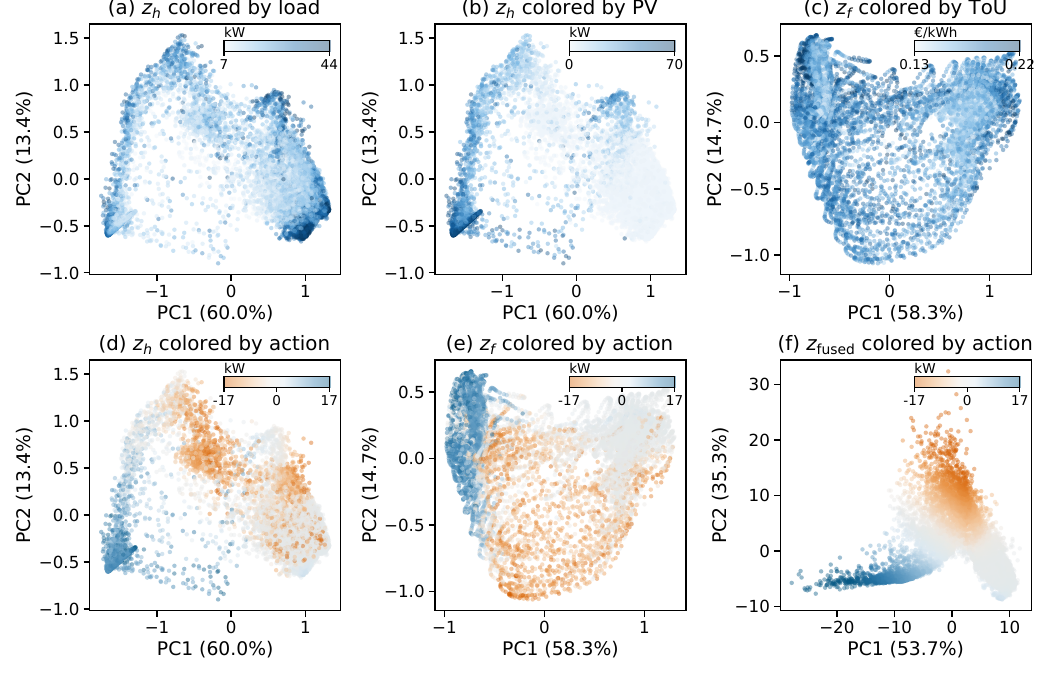}
    \caption{Principal component projection of the R2D\_2TFE latent activations for the policy trained on site ID~12, recorded over the test year at hourly stride (8{,}736 samples). Top row: the historical embedding $\mathbf{z}^{h}_{t}$ colored by load (a) and \gls{pv} generation (b), and the future embedding $\mathbf{z}^{f}_{t}$ colored by the \gls{tou} tariff (c). Bottom row: the same two embeddings (d, e) and the fused representation (f), colored by the raw network setpoint before SOC clipping and thermal derating, on a color scale symmetric about 0~kW. Axis labels give the explained variance of each component; panels sharing a latent space share axis limits.}
    \label{fig:pca}
\end{figure}

To examine what the shared latent representation encodes, the latent activations of the trained R2D\_2TFE policy are recorded during the test rollout and projected by \gls{pca}, which is a post processing step of the benchmark results rather than a separate experiment. Fig.~\ref{fig:pca} shows the projection onto the leading principal components. Both embeddings organize along the exogenous conditions although no input variable is a training target, and the raw network setpoint separates most clearly in the fused representation. Predicting this setpoint linearly from the historical branch alone yields $R^2=0.79$ and from the future branch alone $R^2=0.57$, whereas both branches together yield $R^2=0.90$, an ordering that holds at all five sites. The decision relevant temporal structure is therefore learned implicitly through the behavior cloning objective and is carried by the fusion of both branches rather than by either encoder alone.

\section{Discussion}
\label{sec:discussion}

Building on the results presented above, this section discusses the advantages of R2D drawn from the benchmark study (Section~\ref{sec:results_kpi}), the practical implications drawn from the sensitivity study (Section~\ref{sec:sensitivity_results}) and the \gls{ood} test (Section~\ref{sec:results_ood}), and the limitations of the current study.
 
\subsection{Advantages of R2D}

The advantages discussed here follow from the benchmark study: the annual KPI comparison across the five sites (Fig.~\ref{fig:2kpi}) and the representative weekly operation view (Fig.~\ref{fig:op_profile}). \gls{r2d} exhibits robustness in two complementary senses. First, it is robustly trainable: across all five industrial sites with heterogeneous load profiles, PV generation patterns, and price dynamics, the framework consistently converges to a well-performing policy without site-specific tuning. Second, it robustly clones the behavior of the global expert, inheriting not only the economic performance of the \gls{milp} teacher but also its battery-conservative dispatch behavior. Because the expert itself possesses a detailed system model, including aging dynamics, inverter losses, and thermal constraints, this knowledge is implicitly transferred to \gls{r2d} via the optimizer's output schedule, which serves as the training labels. Crucially, this transfer does not require the student policy to explicitly model the system; the behavioral signal alone is sufficient. As a result, even at sites where the global expert does not achieve top absolute profitability, the learned policy still inherits the battery-conscious dispatch characteristics of the \gls{milp} expert. This battery conservative cloning is visible in the weekly operation view (Fig.~\ref{fig:op_profile}), where R2D tracks the expert dispatch, even though the expert plans with clairvoyant knowledge while R2D observes only a limited window and, where similar observations can carry different expert actions (Fig.~\ref{fig:expert_0_400}, mid-day on Oct.~22nd), converges to an averaged response.
 
Compared to the \gls{mpc} baselines, \gls{r2d} performs favorably across all evaluated sites (Fig.~\ref{fig:2kpi}). The \gls{mpc} benchmarks are more sensitive to forecast quality, and in some configurations, particularly \gls{mpc}\_ML, performance can fall below the persistent baseline, reflecting the compounding of forecast errors within the optimization pipeline. \gls{r2d}, by contrast, bypasses the explicit load and PV forecast--optimization decoupling and is therefore more resilient to such degradation. Furthermore, \gls{r2d} occasionally surpasses even the \gls{mpc}\_CF upper bound. This can be attributed to the structural limitation of rolling-horizon \gls{mpc}: because the optimizer plans over a fixed daily horizon and requires additional end-of-horizon constraints to prevent greedy, short-sighted scheduling, its strategy is sensitive to the choice of terminal conditions. Although the terminal constraint can be relaxed and the \gls{mpc} prediction horizon extended, this introduces greater uncertainty and increases design complexity. In contrast, \gls{r2d} is not subject to horizon-boundary artifacts, as it learns a reactive policy through short-horizon single-step decisions without requiring full-horizon optimization at inference time, while its target single-step actions still retain long-term knowledge. As detailed in Appendix~\ref{sec:app_runtime}, the default \gls{r2d} configuration (1TFE, LSTM-S) requires 20\,355 trainable parameters, 127~MB of initialization memory and a mean per-step decision time of 0.41~ms, compared with 6\,019 parameters, 126~MB and 0.20~ms for the \gls{rl} baseline, and 17~MB and 187--215~ms for the rolling-horizon \gls{mpc} solvers, which hold no trainable parameters. The initialization memory of the learning-based controllers reflects the PyTorch runtime of our development environment and not the size of the policy.

\gls{rl} is also effective under uncertainty, as it can learn control policies with frequent charging and discharging behavior. As shown in Fig.~\ref{fig:2kpi}, even with single-step observations, \gls{rl} reaches 47--72\% of the global optimum. However, such short-sighted greedy charging generally increases battery degradation and leads to high SoH loss across most industrial sites. Site~ID~12 is an exception because battery operation is highly profitable: both the MILP teacher and \gls{r2d} show higher SoH loss than \gls{rl}, as frequent charging and discharging are economically justified by the optimization objective. \gls{r2d} successfully captures and realizes this profitability, whereas \gls{rl} performs fewer actions, resulting in lower aging but also lower revenue. Nevertheless, \gls{r2d} cannot be as compact or fast as \gls{rl} because of its TFE module.

Overall, the benchmark study shows that R2D combines the economic strength and battery conservative behavior of the MILP expert with resilience to forecast error and fast inference, outperforming the tested MPC and RL baselines across all five sites, while remaining below the full-horizon clairvoyant optimum.

\subsection{Impact of Input Profile Characteristics}
\label{sec:discussion_input_profile}

The input profile characteristics influence the entire R2D framework, from the clairvoyant optimum and expert behavior to the final learned policy. Comparing Fig.~\ref{fig:data} and Fig.~\ref{fig:2kpi}, the tested sites can be divided into high-profitability sites, namely IDs~11, 12, and~70 with Global\_CF savings of more than 3000€, and low-profitability sites, namely IDs~10 and~13 with savings of less than 2000~€.

For low-profitability sites, daytime consumption is higher and more frequent, so a larger share of \gls{pv} generation is directly self-consumed. This leaves less flexibility for battery optimization, and additional cycling can lead to degradation costs that partly offset the economic benefit. As shown in Fig.~\ref{fig:expert1213}, the expert behavior in such cases is highly sensitive to the battery-price coefficient: without aging consideration, the optimized solution achieves 2058~€ savings with 4\% \gls{soh} loss, whereas aging-aware optimization reduces \gls{soh} loss to 2.5\% at the cost of 25\% lower savings. A similar trade-off is observed for R2D, with approximately 30\% lower savings when degradation is reduced.

For high-profitability sites, this trade-off is less severe. In ID~12, including aging in the expert objective reduces expert savings by only 4\%, while improving \gls{soh} by 1.3\%. For R2D, the aging-aware expert even improves the learned policy, increasing savings by up to 7\% while also improving \gls{soh} by 1.3\%. This suggests that aging-aware expert demonstrations can provide smoother and more learnable control trajectories when sufficient optimization potential exists.

The profile characteristics also affect robustness under distribution shift. As shown in Fig.~\ref{fig:OOD}, the two effects separate clearly. On its own site, R2D is robust: substituting the load shape, the \gls{pv} profile, or the day-ahead tariff leaves the policy trained on an informative profile close to its in-distribution level, and the tariff can be exchanged almost without loss for either policy. Across sites, in contrast, transferability depends on the training profile: the policy trained on the high-profitability profile holds its level on all five test profiles, whereas the policy trained on the low-profitability profile does not, and degrades most strongly when the \gls{pv} input changes. This is likely because high-profitability profiles contain richer and more diverse battery actions, whereas low-profitability profiles include more idle operation and fewer informative charging or discharging examples. We therefore do not claim generalization to arbitrary unseen conditions, but conclude that a transferable policy can be obtained when the training profile is sufficiently informative, and recommend on this empirical basis selecting a high-profitability profile as the training site when a policy is intended for reuse across sites.

Overall, high-profitability profiles provide more informative demonstrations for training transferable R2D policies, while low-profitability profiles require a more careful choice of expert objective due to the stronger trade-off between economic savings and battery degradation.

\subsection{Practical Implementation Considerations}
\label{sec:discussion_deployment}

For practical implementation, the LSTM backbone is recommended (Fig.~\ref{fig:ablation}) as a robust default architecture, offering stable performance with minimal site-specific tuning. TCN and Transformer models can also be effective, but their performance is more profile-dependent and may require additional architectural adaptation.

The horizon configuration should be kept simple. Historical and forecast windows between one day and one week are computationally feasible, but their impact on performance is limited. In contrast, the control horizon is critical: single-step receding control is preferred for stable and responsive operation (horizon sensitivity, Section~\ref{sec:results_horizon} and Appendix~\ref{sec:app_Horizon}).

The expert model should include as much system knowledge as practically available, since aging- and thermal-aware optimization produces smoother and more learnable demonstrations (Fig.~\ref{fig:expert1213} and Fig.~\ref{fig:expert_0_400}). However, simplified experts remain useful when detailed models are unavailable; even an LP-based expert can provide profitable control behavior for R2D training.

Overall, an industrial implementation should prioritize a robust LSTM-based policy, single-step control execution, and the best feasible expert model rather than excessive architecture or horizon complexity.

\subsection{Limitations}

R2D demonstrates that the representation-to-decision idea works for battery \gls{ems}, and the limitations of this study concern its scope and its implementation rather than the concept. In scope, generalization is bounded: each R2D instance is trained for one profile, the \gls{ood} results in Section~\ref{sec:results_ood} quantify how far such an instance carries beyond its training profile rather than claim general validity, and since all five profiles share the same day-ahead tariff structure, transfer to fundamentally different market designs is untested, as is transfer across battery chemistries, capacities and states of health. Removing the explicit forecasting stage moreover does not remove the underlying uncertainty; it redistributes it into three residual channels: representation error in the \gls{tfe}s, imitation error with respect to the perfect-foresight expert, and partial observability of the future at inference time. The implementation, in turn, is deliberately basic, and its temporal and system awareness can be advanced along the architecture, the expert and the training scheme.

The \gls{lstm}, \gls{tcn} and Transformer backbones evaluated here are standard architectures, and the limited sensitivity to horizon length partly reflects the regular temporal patterns of the EMSx dataset. Richer representations could be obtained by using time-series foundation models as pretrained feature extractors, or architectures such as the Temporal Fusion Transformer that admit static metadata and known future covariates alongside the observed series and thereby improve transferability across sites, together with denoising modules for noisy measurements~\cite{Fan2026LLM, Fan2026EPformer}, within the computational budget available for an industrial \gls{ems} controller.

The teacher is likewise a linearized model, and the performance of the student is bounded by the quality of its demonstrations. Higher-fidelity experts such as nonlinear or dynamic programming solvers, co-optimization frameworks~\cite{Lei2023CoOpt}, or joint EMS--BMS and multi-string formulations with more learnable aging behavior are therefore a natural next step~\cite{Lei2024PINN, Lei2025TCNN}.

The training scheme, finally, is purely offline and provides no explicit constraint handling: over the test year the \gls{soc} limit is binding in 16\% of control steps against 10\% for the clairvoyant expert and 18\% for the receding-horizon \gls{mpc} benchmark; the policy can command discharge at an empty battery (Fig.~\ref{fig:op_profile}, end of the shown week); and the objective weights all action errors equally although an error at a high tariff hour costs more than one at a low tariff hour and an error beyond the \gls{soc} limits is absorbed by the clipping rather than realized. Moving to online and continual learning, fine-tuning the cloned policy with reinforcement learning~\cite{yin2025boosting} (training the temporal policy from scratch with \gls{ppo} does not converge under the current setup, Appendix~\ref{sec:app_rl}), adding physics-informed penalties to the loss~\cite{Lei2024PINN}, and advanced imitation learning such as DAgger, GAIL or IRL are the natural routes to address this.

\section{Conclusion}
\label{sec:conclusion}

This paper introduced \gls{r2d}, an end-to-end imitation learning framework for battery \gls{ems} that directly maps heterogeneous multi-horizon time-series inputs to control decisions through modular \gls{tfe}s, trained via \gls{bc} from an aging-aware \gls{milp} expert. Unlike conventional \gls{mpc} pipelines that accumulate errors across separate forecasting and optimization modules, or standard \gls{rl} that relies on snapshot observations with limited temporal context, \gls{r2d} embeds predictive structure implicitly within latent representations and makes control decisions directly from rich historical and price signals. Benchmarked across five industrial sites against six controllers spanning clairvoyant optimization, forecast-driven \gls{mpc} and an \gls{rl} baseline, \gls{r2d} achieves up to 77\% of the global clairvoyant optimum and delivers 9--28\% higher cost savings than the tested deterministic \gls{mpc} approaches, while producing smoother trajectories and reduced battery degradation inherited from the battery-aware expert.

Ablation studies confirm that \gls{lstm} with a single 15-minute control step is the most robust and deployment-ready backbone, with larger architectures and extended control horizons offering no consistent benefit. Cross-site and single-factor \gls{ood} tests further show that models trained on high-activity profiles generalize well to unseen sites, while models trained on low-activity profiles transfer less consistently, so that \gls{r2d} is a scalable and data-efficient alternative to traditional \gls{ems} approaches when the training profile is sufficiently informative. The full framework is released as open-source Git repository to support reproducibility and future research.

\section*{Acknowledgment}

This work was supported by the Federal Ministry for Economic Affairs and Climate Action (BMWK), Germany, through the project \emph{Storage MultiApp} (FKZ 03EI6081A--E), managed by Projektträger Jülich; by Interreg VI Alpenrhein-Bodensee-Hochrhein and the European Union through the project \emph{ENABLE} (ABH046); and by the Bayerische Transfo-rmations- und Forschungsstiftung through the project \emph{KIM-Bat} (AZ-1563-22).

\section*{Declaration of competing interests}
The authors declare that they have no known competing financial interests or personal relationships that could have appeared to influence the work reported in this paper.

\printcredits

\section*{Declaration of generative AI use}
During the preparation of this work, the authors used CodeX and Claude Code to improve the language and readability of the document. After using this tool/service, the authors reviewed and edited the content as needed and take full responsibility for the content of the published article.

\section*{Data availability}
The raw data used in this research is discussed in Section \ref{sec:data}, and the processed data is available in the proposed Git repository \texttt{R2D}~\cite{R2D2026}.

\bibliographystyle{cas-model2-names}
\bibliography{Bib}

\appendix

\section{BESS simulation model} \label{sec:BESS sim}
The detailed \gls{bess} simulation presented in Fig.~\ref{fig:usecase} is explained in this section. The simulation model updates the battery state from the AC-side battery command over a fixed time step $\Delta t$. To align with the notation used in the paper, the AC-side charge and
discharge powers are denoted by $p_t^{B,ch}$ and $p_t^{B,dch}$, and the net AC-side battery power
is written as
\begin{equation}
p_t^B = p_t^{B,ch} - p_t^{B,dch}.
\end{equation}
The available battery capacity decreases with state of health, so the effective usable energy is
\begin{equation}
E_t^{N,avail} = soh_t E^N .
\end{equation}

\begin{figure}[t]
    \centering
    \includegraphics[width=0.48\textwidth]{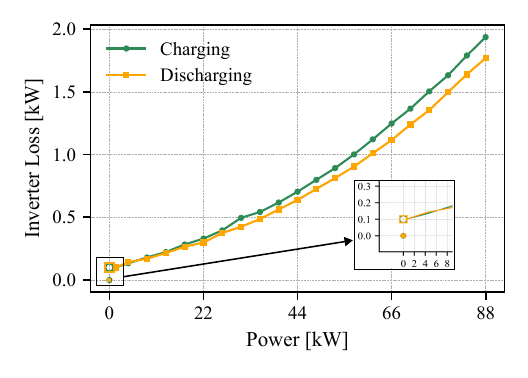}
    \caption{Piece-wise linear approximation of the experimentally acquired inverter loss function of the industrial inverter used in this study~\cite{Tanjavooru.2025}.}
    \label{fig:inverter}
\end{figure}

The inverter loss is represented by one-dimensional interpolation of charge and discharge lookup
tables as shown in Fig.~\ref{fig:inverter},
\begin{equation}
p_t^{inv} =
\begin{cases}
\operatorname{interp}\!\left(|p_t^B|;\mathcal{P}^{ch},\mathcal{L}^{ch}\right), & p_t^B > 0,\\[4pt]
\operatorname{interp}\!\left(|p_t^B|;\mathcal{P}^{dch},\mathcal{L}^{dch}\right), & p_t^B < 0,\\[4pt]
0, & p_t^B = 0,
\end{cases}
\end{equation}
so that the DC-side battery power before clipping is
\begin{equation}
p_t^{B,dc} = p_t^B - p_t^{inv}.
\end{equation}

The admissible charge and discharge powers are constrained by the current state of charge,
\begin{align}
p_t^{B,ch,\max} &= \frac{(soc_{\max}-soc_t)E_t^{N,avail}}{\Delta t},\\
p_t^{B,dch,\max} &= \frac{(soc_t-soc_{\min})E_t^{N,avail}}{\Delta t},
\end{align}
and the clipped DC-side power is
\begin{equation}
p_t^{B,dc,clip} =
\begin{cases}
\min\!\left(p_t^{B,dc},p_t^{B,ch,\max}\right), & p_t^{B,dc}>0,\\[4pt]
\max\!\left(p_t^{B,dc},-p_t^{B,dch,\max}\right), & p_t^{B,dc}\le 0.
\end{cases}
\end{equation}
Thermal derating is then applied through a temperature-dependent factor to constrain the battery operation in the optimal operating temperature range~\cite{Tanjavooru.2026},
\begin{equation}
p_t^{B,dc,der} = d(T_t)\, p_t^{B,dc,clip}.
\end{equation}

\begin{figure}[t]
    \centering
    \includegraphics[width=0.48\textwidth]{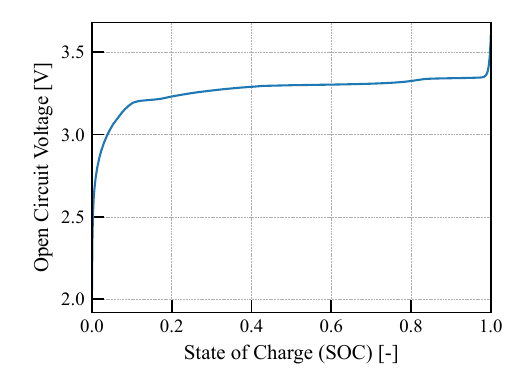}
    \caption{SOC--OCV relationship for \gls{lfp} cells derived from literature~\cite{Moller.2022b}.}
    \label{fig:soc_ocv}
\end{figure}

The open-circuit voltage is interpolated from SOC--OCV relationship for \gls{lfp} cells as shown in Fig.~\ref{fig:soc_ocv},
\begin{equation}
v_t^{oc} =
\operatorname{interp}\!\left(soc_t;\mathcal{SOC},\mathcal{OCV}\right)
\end{equation}

The \gls{bess} consists of four identical battery strings connected in parallel, each modeled as a 130-series-cell electro-thermal unit, resulting in an aggregated 0D system representation. Using the implemented configuration with $N_s=130$ and $N_p=4$, the \gls{bess} is represented as a single aggregated 0D electro-thermal model. The terminal voltage and current are approximated by
\begin{equation}
v_t =
\begin{cases}
v_t^{oc} + \dfrac{|p_t^{B,dc,der}|}{N_p N_s 10^{-3} v_t^{oc}} R, & p_t^B > 0,\\[10pt]
v_t^{oc} - \dfrac{|p_t^{B,dc,der}|}{N_p N_s 10^{-3} v_t^{oc}} R, & p_t^B \le 0,
\end{cases}
\end{equation}

\begin{equation}
i_t^B = \frac{v_t - v_t^{oc}}{R}.
\end{equation}

Heat generation and heat dissipation per unit area are modeled at the system level as
\begin{align}
q_t^{gen} &= \frac{(i_t^B)^2 R N_s N_p}{A},\\
q_t^{diss} &= h(T_t - T_t^{air}),
\end{align}

The battery temperature is updated using a lumped 0D thermal model based on the net heat flux:
\begin{equation}
T_{t+1} = T_t + \frac{(q_t^{gen} - q_t^{diss}) \Delta t \cdot 3600}{\rho H c_p}.
\end{equation}

The state of charge is then updated from the derated DC-side battery power after subtracting the
heat proxy,
\begin{equation}
soc_{t+1} = soc_t + \frac{(p_t^{B,dc,der}-p_t^{heat})\Delta t}{E_t^{N,avail}}.
\end{equation}
where, 
\begin{equation}
p_t^{heat} = 10^{-3} q_t^{gen}.
\end{equation}

The aging formulation used in this work is derived from the comprehensive electro-thermal battery aging model proposed in~\cite{schimpe2018comprehensive}. To improve tractability for optimization, auxiliary temperature-, voltage-, current-, and throughput-dependent factors are introduced. In the following, LT, HT, and HSOC denote low-temperature, high-temperature, and high-state-of-charge operating regions, respectively. To shorten the aging equations, auxiliary temperature, voltage, and throughput factors are introduced. 

The calendar-aging temperature and voltage factors are defined as
\begin{equation}
\phi_t^{T,cal}
= \exp\!\left[
-\frac{E_{a,\mathrm{cal}}}{R_g}
\left(
\frac{1}{T_t+273} - \frac{1}{T_{\mathrm{ref}}}
\right)
\right],
\end{equation}

\begin{equation}
\phi_t^{V}
= \exp\!\left[
\frac{\alpha F}{R_g}\frac{U_{a,\mathrm{ref}}-v_t^{oc}}{T_{\mathrm{ref}}}
\right] + k_0,
\end{equation}

and the calendar throughput term is
\begin{equation}
\psi_t^{cal} = \sqrt{\tau_t+\Delta t}-\sqrt{\tau_t}.
\end{equation}

The calendar-aging increment is then written as
\begin{equation}
\Delta soh_t^{cal}
= 0.25\,k_{\mathrm{cal,ref}} \, \phi_t^{T,cal} \, \phi_t^{V} \, \psi_t^{cal}.
\end{equation}

For cycling aging, the high-temperature, HT stress factor and throughput term are defined as
\begin{equation}
\phi_t^{T,HT}
= \exp\!\left[
-\frac{E_{a,\mathrm{HT}}}{R_g}
\left(
\frac{1}{T_t}-\frac{1}{T_{\mathrm{ref}}}
\right)
\right],
\end{equation}

\begin{equation}
\psi_t^{cyc}
= \sqrt{|i_t^B|(\tau_t+\Delta t)}-\sqrt{|i_t^B|\tau_t}.
\end{equation}

The high-temperature cycling increment is
\begin{equation}
\Delta soh_t^{HT}
= k_{\mathrm{HT,ref}} \, \phi_t^{T,HT} \, \psi_t^{cyc}.
\end{equation}

The low-temperature, LT stress factors are
\begin{equation}
\phi_t^{T,LT}
= \exp\!\left[
-\frac{E_{a,\mathrm{LT}}}{R_g}
\left(
\frac{1}{T_t}-\frac{1}{T_{\mathrm{ref}}}
\right)
\right],
\end{equation}

\begin{equation}
\phi_t^{I,LT}
= \exp\!\left[
\beta_{\mathrm{LT}}\frac{i_t^B-I_{\mathrm{ch,ref}}}{C_0}
\right].
\end{equation}

The low-temperature cycling increment is
\begin{equation}
\Delta soh_t^{LT}
= \mathbb{I}(i_t^B>0)\,
k_{\mathrm{LT,ref}} \, \phi_t^{T,LT} \, \phi_t^{I,LT} \, \psi_t^{cyc}.
\end{equation}

For the low-temperature high-SOC, LT–HSOC regime, the temperature factor, current factor, and throughput term are defined as
\begin{equation}
\phi_t^{T,LT\text{-}HSOC}
= \exp\!\left[
\frac{E_{a,\mathrm{LT\text{-}HSOC}}}{R_g}
\left(
\frac{1}{T_{\mathrm{ref}}}-\frac{1}{T_t}
\right)
\right],
\end{equation}

\begin{equation}
\chi_t^{HSOC}
= \frac{\operatorname{sign}(soc_t-SOC_{\mathrm{ref}})+1}{2},
\end{equation}

\begin{equation}
\phi_t^{I,LT\text{-}HSOC}
= \exp\!\left[
\beta_{\mathrm{LT\text{-}HSOC}}
\left(\frac{i_t^B-I_{\mathrm{ch,ref}}}{C_0}\right)
\chi_t^{HSOC}
\right],
\end{equation}

\begin{equation}
\psi_t^{HSOC}
= |i_t^B|(\tau_t+\Delta t)-|i_t^B|\tau_t.
\end{equation}

\begin{equation}
\kappa_t^{LT\text{-}HSOC}
= \mathbb{I}(i_t^B>0)\,k_{\mathrm{LT\text{-}HSOC,ref}}.
\end{equation}

The LT–HSOC charging increment is
\begin{equation}
\Delta soh_t^{LT\text{-}HSOC} =
\kappa_t^{LT\text{-}HSOC} \,
\phi_t^{T,LT\text{-}HSOC} \,
\phi_t^{I,LT\text{-}HSOC} \,
\psi_t^{HSOC}.
\end{equation}

Finally, the total state-of-health evolution is given by
\begin{equation}
soh_{t+1}
= soh_t
- \Delta soh_t^{cal}
- \Delta soh_t^{HT}
- \Delta soh_t^{LT}
- \Delta soh_t^{LT\text{-}HSOC}.
\end{equation}

\section{Constant values}
\label{sec:values}
The following table (Table~\ref{tab:bess_params}) summarizes the BESS thermal modeling and aging estimation parameters used throughout this study.
\begin{table}[t]
\centering
\caption{\textbf{{\gls{bess} thermal modeling and aging estimation parameters used in the study.}}}
\label{tab:bess_params}
\begin{tabular}{>{\raggedright\arraybackslash}p{3.8cm} >{\raggedright\arraybackslash}p{3.0cm}}
\toprule
Constant & Value \\
\midrule
$\Delta t$ & $0.25$ h \\
$E^N$ & $100$ kWh \\
$soc_{\min}$ & $0.1$ \\
$soc_{\max}$ & $0.9$ \\
$soc_{start}$ & $0.1$ \\
$soh_{start}$ & $1.0$ \\
$T_{start}$ & $25\,^{\circ}\mathrm{C}$ \\
$T_t^{air}$ & $25\,^{\circ}\mathrm{C}$ \\
$R$ & $0.00225\,\Omega$ \\
$A$ & $2.1248$ m$^2$ \\
$h$ & $16$ W\,m$^{-2}$\,K$^{-1}$ \\
$\rho$ & $3300$ kg\,m$^{-3}$ \\
$c_p$ & $1258$ J\,kg$^{-1}$\,K$^{-1}$ \\
$H$ & $0.174$ m \\
$k_{\mathrm{cal,ref}}$ & $3.694\times10^{-4}$ \\
$E_{a,\mathrm{cal}}$ & $20592$ J\,mol$^{-1}$ \\
$R_g$ & $8.314$ J\,mol$^{-1}$\,K$^{-1}$ \\
$\alpha$ & $0.384$ \\
$F$ & $96485$ C\,mol$^{-1}$ \\
$U_{a,\mathrm{ref}}$ & $0.123$ V \\
$T_{\mathrm{ref}}$ & $298.15$ K \\
$k_0$ & $0.142$ \\
$k_{\mathrm{HT,ref}}$ & $1.456\times10^{-4}$ \\
$E_{a,\mathrm{HT}}$ & $32699$ J\,mol$^{-1}$ \\
$k_{\mathrm{LT,ref}}$ & $4.009\times10^{-4}$ \\
$E_{a,\mathrm{LT}}$ & $55546$ J\,mol$^{-1}$ \\
$\beta_{\mathrm{LT}}$ & $2.64$ \\
$I_{\mathrm{ch,ref}}$ & $3.0$ A \\
$C_0$ & $3.0$ Ah \\
$SOC_{\mathrm{ref}}$ & $0.82$ \\
$k_{\mathrm{LT\text{-}HSOC,ref}}$ & $2.031\times10^{-6}$ \\
$E_{a,\mathrm{LT\text{-}HSOC}}$ & $2.3\times10^5$ J\,mol$^{-1}$ \\
$\beta_{\mathrm{LT\text{-}HSOC}}$ & $7.8$ \\ 
$c_{cal}^{T}$          & $4.92\times10^{-7}\rho^{B}$ \\
$c_{cal}^{soc}$        & $0$ \\
$c_{cyc}^{ch}$         & $1.29\times10^{-4}\rho^{B}$ \\
$c_{cyc}^{dch}$        & $1.30\times10^{-4}\rho^{B}$ \\
\bottomrule
\end{tabular}
\end{table}

\section{Optimization Models}
\label{app:opt_models}

The 1D-MILP expert, described in Section~\ref{sec:demo}, serves as the
reference formulation with objective
\begin{equation}
  \min\;J^{\mathrm{1D\text{-}MILP}} = \mathbb{C}^{E} + \mathbb{C}^{B,cal} + \mathbb{C}^{B,cyc},
\end{equation}
where $\mathbb{C}^{B,cal}$ captures both temperature- and SoC-dependent calendar
aging and $\mathbb{C}^{B,cyc}$ penalises charge/discharge throughput. The two
benchmark experts simplify this progressively. The LP expert minimises
\begin{equation}
  \min\;J^{\mathrm{LP}} = \mathbb{C}^{E}
\end{equation}
alone, dropping all aging penalties and therefore requiring no binary variables
or thermal dynamics. The aging-aware MILP expert adopts the same three-term
structure,
\begin{equation}
  \min\;J^{\mathrm{MILP}} = \mathbb{C}^{E} + \mathbb{C}^{B,cal} + \mathbb{C}^{B,cyc},
\end{equation}
but simplifies $\mathbb{C}^{B,cal}$ to a base rate plus a high-SoC excess
penalty, omitting the temperature dependence and thus removing the need for a
thermal model and inverter-loss constraints.

The stochastic expert extends the 1D-MILP by replacing $\mathbb{C}^{E}$
with a scenario-averaged electricity cost $\mathbb{C}^{E,SQ} =
\mathbb{E}_{q\in Q}[J_q^{\mathrm{energy}}]$ over paired quantile scenarios
$q\in Q$, giving
\begin{equation}
  \min\;J^{\mathrm{SQ}} = \mathbb{C}^{E,SQ} + \mathbb{C}^{B,cal} + \mathbb{C}^{B,cyc}.
\end{equation}
Battery decisions are shared across all scenarios, so the aging costs are
evaluated on a single trajectory; only the grid exchange variables become
scenario-specific, introducing per-scenario AC balance and grid exclusivity
constraints.

In all aging-aware formulations, the degradation coefficients within
$\mathbb{C}^{B,cal}$ and $\mathbb{C}^{B,cyc}$ are scaled by the battery
investment value $\rho^B$, set to 400~€/kWh as the benchmark scenario,
to control the trade-off between electricity cost and battery longevity.
No such parameter applies to the LP expert.

\section{Full results}
\label{sec:full_results}
This appendix provides the complete numeric results underlying Fig.~\ref{fig:2kpi}. Table~\ref{tab:full_results} reports the mean and standard deviation of normalized savings $\bar{S}^{B}$, annual \gls{soh} loss $\Delta$SOH, and battery efficiency $\eta$ for all controllers across all five industrial sites.

\begin{table*}[t]
\centering
\scriptsize
\caption{Final benchmark numeric summary. Rows with multiple seeds are reported as mean $\pm$ standard deviation.}
\label{tab:full_results}
\begin{tabular}{rlrrrrr}
\toprule
ID & Model type & Saving (€) & $\bar{S}^{Global_CF}$ (\%) & Saving / MPC CF (\%) & $\Delta$ SOH (\%) & Efficiency (\%) \\
\midrule
10 & Global\_CF & 1834.8 & 100.0 & 104.5 & 1.603 & 89.10 \\
10 & MPC\_CF & 1755.1 & 95.7 & 100.0 & 1.576 & 89.24 \\
10 & R2D\_1TFE & $1138.1 \pm 52.4$ & $62.0 \pm 2.9$ & $64.8 \pm 3.0$ & $1.700 \pm 0.097$ & $87.99 \pm 0.50$ \\
10 & R2D\_2TFE & $1151.1 \pm 81.3$ & $62.7 \pm 4.4$ & $65.6 \pm 4.6$ & $1.684 \pm 0.096$ & $87.93 \pm 0.66$ \\
10 & MPC\_ML & $680.0 \pm 205.9$ & $37.1 \pm 11.2$ & $38.7 \pm 11.7$ & $1.313 \pm 0.329$ & $86.99 \pm 4.20$ \\
10 & RL\_STD & $921.1 \pm 136.0$ & $50.2 \pm 7.4$ & $52.5 \pm 7.8$ & $2.430 \pm 0.319$ & $89.74 \pm 0.84$ \\
10 & MPC\_PQ & 993.4 & 54.1 & 56.6 & 1.771 & 90.22 \\
10 & MPC\_PD & 820.7 & 44.7 & 46.8 & 1.616 & 89.26 \\
\midrule
11 & Global\_CF & 3426.8 & 100.0 & 147.7 & 2.969 & 93.12 \\
11 & MPC\_CF & 2320.5 & 67.7 & 100.0 & 2.373 & 91.67 \\
11 & R2D\_1TFE & $2387.6 \pm 80.5$ & $69.7 \pm 2.4$ & $102.9 \pm 3.5$ & $2.711 \pm 0.090$ & $92.25 \pm 0.22$ \\
11 & R2D\_2TFE & $2372.4 \pm 53.7$ & $69.2 \pm 1.6$ & $102.2 \pm 2.3$ & $2.733 \pm 0.065$ & $92.29 \pm 0.14$ \\
11 & MPC\_ML & $1134.2 \pm 83.9$ & $33.1 \pm 2.4$ & $48.9 \pm 3.6$ & $2.299 \pm 0.108$ & $91.00 \pm 0.48$ \\
11 & RL\_STD & $1980.2 \pm 117.5$ & $57.8 \pm 3.4$ & $85.3 \pm 5.1$ & $2.941 \pm 0.157$ & $92.32 \pm 0.38$ \\
11 & MPC\_PQ & 1468.8 & 42.9 & 63.3 & 2.489 & 92.18 \\
11 & MPC\_PD & 1382.2 & 40.3 & 59.6 & 2.561 & 92.42 \\
\midrule
12 & Global\_CF & 4283.4 & 100.0 & 111.9 & 3.167 & 93.69 \\
12 & MPC\_CF & 3827.4 & 89.4 & 100.0 & 3.047 & 93.45 \\
12 & R2D\_1TFE & $3299.9 \pm 99.4$ & $77.0 \pm 2.3$ & $86.2 \pm 2.6$ & $3.075 \pm 0.093$ & $93.40 \pm 0.12$ \\
12 & R2D\_2TFE & $3249.7 \pm 98.1$ & $75.9 \pm 2.3$ & $84.9 \pm 2.6$ & $3.028 \pm 0.095$ & $93.32 \pm 0.13$ \\
12 & MPC\_ML & $2758.2 \pm 51.0$ & $64.4 \pm 1.2$ & $72.1 \pm 1.3$ & $3.450 \pm 0.126$ & $94.15 \pm 0.10$ \\
12 & RL\_STD & $3088.3 \pm 66.8$ & $72.1 \pm 1.6$ & $80.7 \pm 1.7$ & $2.955 \pm 0.124$ & $92.42 \pm 0.44$ \\
12 & MPC\_PQ & 2870.1 & 67.0 & 75.0 & 3.193 & 93.76 \\
12 & MPC\_PD & 2634.1 & 61.5 & 68.8 & 3.204 & 93.79 \\
\midrule
13 & Global\_CF & 1512.3 & 100.0 & 108.4 & 1.485 & 88.46 \\
13 & MPC\_CF & 1394.6 & 92.2 & 100.0 & 1.436 & 88.74 \\
13 & R2D\_1TFE & $933.0 \pm 63.0$ & $61.7 \pm 4.2$ & $66.9 \pm 4.5$ & $1.499 \pm 0.043$ & $86.17 \pm 0.59$ \\
13 & R2D\_2TFE & $915.9 \pm 47.2$ & $60.6 \pm 3.1$ & $65.7 \pm 3.4$ & $1.511 \pm 0.057$ & $86.33 \pm 0.42$ \\
13 & MPC\_ML & $757.6 \pm 15.6$ & $50.1 \pm 1.0$ & $54.3 \pm 1.1$ & $1.225 \pm 0.024$ & $87.29 \pm 0.25$ \\
13 & RL\_STD & $714.6 \pm 82.5$ & $47.3 \pm 5.5$ & $51.2 \pm 5.9$ & $1.674 \pm 0.136$ & $87.16 \pm 0.85$ \\
13 & MPC\_PQ & 517.5 & 34.2 & 37.1 & 1.394 & 88.68 \\
13 & MPC\_PD & 515.8 & 34.1 & 37.0 & 1.394 & 88.68 \\
\midrule
70 & Global\_CF & 3217.1 & 100.0 & 137.3 & 3.043 & 93.03 \\
70 & MPC\_CF & 2343.5 & 72.8 & 100.0 & 2.468 & 91.34 \\
70 & R2D\_1TFE & $2318.1 \pm 43.7$ & $72.1 \pm 1.4$ & $98.9 \pm 1.9$ & $2.881 \pm 0.075$ & $92.48 \pm 0.18$ \\
70 & R2D\_2TFE & $2390.7 \pm 64.7$ & $74.3 \pm 2.0$ & $102.0 \pm 2.8$ & $2.939 \pm 0.071$ & $92.68 \pm 0.17$ \\
70 & MPC\_ML & $1480.1 \pm 37.1$ & $46.0 \pm 1.2$ & $63.2 \pm 1.6$ & $2.539 \pm 0.034$ & $91.49 \pm 0.19$ \\
70 & RL\_STD & $2032.1 \pm 67.7$ & $63.2 \pm 2.1$ & $86.7 \pm 2.9$ & $3.166 \pm 0.143$ & $93.05 \pm 0.31$ \\
70 & MPC\_PQ & 1623.3 & 50.5 & 69.3 & 2.598 & 91.67 \\
70 & MPC\_PD & 1510.7 & 47.0 & 64.5 & 2.556 & 91.63 \\
\bottomrule
\end{tabular}
\end{table*}

\section{Sensitivity Analysis: Horizon}
\label{sec:app_Horizon}
The sensitivity of normalized energy-bill savings to horizon specifications is reported in Tables~\ref{tab:horizon_sensitivity_pagewide}. Foresight horizon has negligible impact across all sites, with savings remaining stable within $\pm$2\% regardless of the chosen horizon. The historical horizon exhibits similarly stable behavior, with no consistent directional effect across sites. In contrast, the control horizon shows a clear degradation with increasing length: savings drop substantially as the horizon extends from 15~min to 24~h, with the effect most pronounced at sites 10 and 12, suggesting that longer control horizons introduce suboptimal scheduling decisions that erode performance regardless of the model specification.
 
\begin{table*}[t]
\centering
\scriptsize
\setlength{\tabcolsep}{3.5pt}
\caption{Horizon sensitivity by site in page-wide format. Entries report mean $\pm$ standard deviation.}
\label{tab:horizon_sensitivity_pagewide}
\resizebox{\textwidth}{!}{%
\begin{tabular}{c*{10}{c}}
\toprule
& \multicolumn{2}{c}{ID 10} & \multicolumn{2}{c}{ID 11} & \multicolumn{2}{c}{ID 12} & \multicolumn{2}{c}{ID 13} & \multicolumn{2}{c}{ID 70} \\
\cmidrule(lr){2-3}\cmidrule(lr){4-5}\cmidrule(lr){6-7}\cmidrule(lr){8-9}\cmidrule(lr){10-11}
Horizon (h) & 1TFE & 2TFE & 1TFE & 2TFE & 1TFE & 2TFE & 1TFE & 2TFE & 1TFE & 2TFE \\
\midrule
\multicolumn{11}{l}{\textbf{Historical horizon}} \\
\midrule
24  & $57.0\pm5.9$ & $59.2\pm3.6$ & $67.3\pm2.6$ & $68.9\pm2.0$ & $74.8\pm3.2$ & $76.6\pm2.3$ & $59.3\pm5.0$ & $61.6\pm1.9$ & $72.9\pm2.3$ & $71.6\pm1.5$ \\
72  & $54.9\pm5.7$ & $59.2\pm3.0$ & $67.3\pm2.9$ & $70.0\pm1.9$ & $76.4\pm3.2$ & $75.6\pm1.9$ & $60.4\pm3.8$ & $61.5\pm2.5$ & $72.3\pm1.9$ & $72.5\pm1.3$ \\
120 & $53.6\pm8.8$ & $57.8\pm5.0$ & $67.2\pm2.5$ & $68.7\pm2.5$ & $75.9\pm2.6$ & $75.0\pm2.2$ & $61.4\pm2.7$ & $61.0\pm2.1$ & $72.7\pm2.7$ & $72.0\pm2.3$ \\
168 & $53.6\pm8.9$ & $59.5\pm4.2$ & $67.5\pm3.0$ & $69.2\pm2.7$ & $75.7\pm2.7$ & $76.2\pm1.0$ & $59.6\pm2.5$ & $61.1\pm2.5$ & $71.9\pm2.1$ & $72.2\pm2.2$ \\
\midrule
\multicolumn{11}{l}{\textbf{Foresight horizon}} \\
\midrule
1  &  & $60.6\pm3.1$ &  & $67.1\pm3.7$ &  & $75.4\pm1.5$ &  & $61.1\pm3.5$ &  & $72.7\pm4.2$ \\
4  &  & $60.2\pm3.5$ &  & $68.0\pm3.0$ &  & $75.0\pm3.2$ &  & $59.9\pm4.6$ &  & $72.7\pm2.8$ \\
8  &  & $57.4\pm4.1$ &  & $67.6\pm2.6$ &  & $74.3\pm3.1$ &  & $61.0\pm2.0$ &  & $71.5\pm4.4$ \\
12 &  & $56.7\pm3.7$ &  & $68.2\pm2.7$ &  & $73.7\pm1.4$ &  & $61.7\pm1.8$ &  & $72.9\pm1.8$ \\
16 &  & $56.4\pm3.8$ &  & $67.3\pm2.9$ &  & $74.1\pm2.5$ &  & $60.4\pm2.8$ &  & $71.4\pm2.0$ \\
20 &  & $57.9\pm6.0$ &  & $69.7\pm2.2$ &  & $74.8\pm2.2$ &  & $60.7\pm3.7$ &  & $73.1\pm1.6$ \\
24 & $55.1\pm6.0$ & $59.1\pm2.9$ & $67.3\pm2.8$ & $69.6\pm1.9$ & $75.3\pm3.3$ & $75.5\pm2.2$ & $59.8\pm4.0$ & $61.1\pm2.8$ & $71.8\pm2.1$ & $72.2\pm1.7$ \\
\midrule
\multicolumn{11}{l}{\textbf{Control horizon}} \\
\midrule
0.25 & $55.0\pm5.9$ & $59.2\pm3.0$ & $67.3\pm2.9$ & $70.0\pm1.9$ & $76.4\pm3.2$ & $75.6\pm1.9$ & $60.4\pm3.8$ & $61.5\pm2.5$ & $72.3\pm1.9$ & $72.5\pm1.3$ \\
1    & $51.6\pm6.1$ & $57.6\pm4.1$ & $62.7\pm2.5$ & $63.3\pm3.2$ & $71.6\pm2.3$ & $71.8\pm2.8$ & $57.7\pm4.4$ & $58.2\pm3.5$ & $71.0\pm1.3$ & $69.9\pm3.5$ \\
4    & $46.5\pm3.3$ & $51.7\pm2.2$ & $53.1\pm2.3$ & $55.7\pm2.1$ & $62.1\pm2.2$ & $62.7\pm2.0$ & $53.5\pm3.6$ & $52.5\pm3.0$ & $65.2\pm1.8$ & $64.0\pm1.2$ \\
8    & $37.0\pm5.1$ & $43.6\pm4.1$ & $51.7\pm1.9$ & $52.5\pm1.1$ & $63.4\pm1.7$ & $60.6\pm2.1$ & $49.0\pm3.3$ & $46.9\pm3.5$ & $61.7\pm1.8$ & $61.0\pm2.2$ \\
12   & $48.0\pm4.5$ & $51.0\pm4.8$ & $54.6\pm2.1$ & $54.8\pm1.5$ & $62.5\pm2.5$ & $60.4\pm1.8$ & $51.7\pm2.2$ & $52.1\pm1.8$ & $61.4\pm2.0$ & $62.0\pm1.2$ \\
16   & $38.5\pm4.9$ & $41.6\pm4.9$ & $51.4\pm2.1$ & $51.8\pm2.9$ & $59.4\pm3.3$ & $57.8\pm3.2$ & $49.0\pm3.1$ & $47.9\pm2.5$ & $59.5\pm1.9$ & $59.5\pm2.0$ \\
20   & $38.5\pm8.5$ & $44.1\pm5.0$ & $52.5\pm1.8$ & $52.8\pm0.9$ & $59.0\pm2.8$ & $60.3\pm1.7$ & $47.5\pm1.9$ & $47.7\pm2.4$ & $60.8\pm1.1$ & $60.4\pm1.9$ \\
24   & $28.0\pm18.4$ & $37.9\pm15.0$ & $52.5\pm1.8$ & $52.6\pm2.5$ & $58.5\pm3.1$ & $58.4\pm2.2$ & $45.9\pm2.6$ & $46.0\pm2.7$ & $60.1\pm2.5$ & $60.1\pm1.7$ \\
\bottomrule
\end{tabular}%
}
\end{table*}

\section{Run Time}
\label{sec:app_runtime}

Table~\ref{tab:runtime-summary-runtime-only} reports the runtime of representative benchmark and R2D controllers selected from the corresponding benchmark and R2D model sets. All controllers were evaluated under an identical execution setting: test-time replay on the same site and horizon, run sequentially on a single CPU core with single-threaded numerical backends to ensure a fair comparison. \textit{Online runtime (s)} denotes the accumulated controller decision time over the full evaluation horizon. \textit{Mean step (s)} and \textit{P95 step (s)} summarize the typical and near-tail latency of individual control decisions. For memory, \textit{Init mem. (MB)} captures the incremental process memory after controller construction, which for the learning-based controllers reflects the PyTorch runtime rather than the policy, whose weights occupy about 81~kB for R2D and 24~kB for the \gls{rl} baseline, while \textit{Max step mem. (MB)} records the largest additional memory increase observed during any single call.

Training time was not recorded under a controlled setting. The policies were trained partly on a shared workstation running Ubuntu~24.04~LTS with an AMD Ryzen Threadripper PRO 5995WX processor (64~cores, 128~threads), 512~GB of system memory and two NVIDIA RTX~A6000 GPUs, concurrently with unrelated jobs of other users, and partly on a MacBook Pro with an Apple M4~Pro processor and 24~GB of unified memory. The following values are therefore indicative of the order of magnitude rather than comparable between configurations, and we report them for completeness; the online runtime in Table~\ref{tab:runtime-summary-runtime-only} is measured under an identical execution setting and is the quantity relevant for deployment. Averaged over ten seeds, training takes 672~s for R2D\_1TFE\_LSTM\_S and 750~s for R2D\_2TFE\_LSTM\_S, and lies between 672 and 4573~s for the LSTM presets, between 1319 and 9489~s for the Transformer presets, and between 1118 and 24227~s for the TCN presets, with early stopping terminating training after 22 to 32 epochs.

\begin{table*}[t]
\centering
\scriptsize
\caption{Online runtime and controller-memory.}
\label{tab:runtime-summary-runtime-only}
\begin{tabular}{lrrrrr}
\toprule
Model & Online runtime (s) & Mean step (s) & P95 step (s) & Init mem. (MB) & Max step mem. (MB) \\
\midrule
Global\_CF & 103.054625 & 103.054625 & 103.054625 & 16.703125 & 5533.289062 \\
MPC\_CF & 68.148713 & 0.186709 & 0.242233 & 16.992188 & 39.109375 \\
MPC\_PD & 78.709414 & 0.215642 & 0.269154 & 16.949219 & 29.531250 \\
MPC\_ML\_LSTM & 70.191954 & 0.192307 & 0.282719 & 141.347656 & 44.562500 \\
MPC\_ML\_TCN & 70.096798 & 0.192046 & 0.282316 & 142.363281 & 41.015625 \\
MPC\_ML\_TRANSFORMER & 75.070992 & 0.205674 & 0.297718 & 148.343750 & 37.207031 \\
MPC\_PQ & 381.448106 & 1.045063 & 1.142026 & 16.066406 & 52.218750 \\
RL\_STD & 6.909897 & 0.000197 & 0.000203 & 126.488281 & 1.460938 \\
R2D\_1TFE\_LSTM\_S & 14.438348 & 0.000412 & 0.000418 & 126.671875 & 6.921875 \\
R2D\_1TFE\_LSTM\_M & 20.595747 & 0.000588 & 0.000594 & 126.648438 & 7.027344 \\
R2D\_1TFE\_LSTM\_L & 21.743704 & 0.000621 & 0.000626 & 126.730469 & 7.144531 \\
R2D\_1TFE\_LSTM\_XL & 23.334781 & 0.000666 & 0.000672 & 126.816406 & 7.152344 \\
R2D\_1TFE\_TCN\_S & 33.754404 & 0.000963 & 0.000973 & 127.097656 & 2.136719 \\
R2D\_1TFE\_TCN\_M & 38.757103 & 0.001106 & 0.001114 & 127.136719 & 2.136719 \\
R2D\_1TFE\_TCN\_L & 45.699420 & 0.001304 & 0.001314 & 127.472656 & 2.136719 \\
R2D\_1TFE\_TCN\_XL & 63.134512 & 0.001802 & 0.001812 & 126.714844 & 2.386719 \\
R2D\_1TFE\_TRANSFORMER\_S & 31.984467 & 0.000913 & 0.000942 & 128.421875 & 2.507812 \\
R2D\_1TFE\_TRANSFORMER\_M & 45.016961 & 0.001285 & 0.001298 & 127.613281 & 2.421875 \\
R2D\_1TFE\_TRANSFORMER\_L & 53.750969 & 0.001534 & 0.001563 & 128.718750 & 2.523438 \\
R2D\_1TFE\_TRANSFORMER\_XL & 74.681722 & 0.002131 & 0.002161 & 127.757812 & 2.988281 \\
R2D\_2TFE\_LSTM\_S & 19.195612 & 0.000548 & 0.000554 & 126.656250 & 7.101562 \\
R2D\_2TFE\_LSTM\_M & 25.806566 & 0.000736 & 0.000743 & 126.824219 & 7.214844 \\
R2D\_2TFE\_LSTM\_L & 27.028607 & 0.000771 & 0.000777 & 126.843750 & 7.257812 \\
R2D\_2TFE\_LSTM\_XL & 31.430737 & 0.000897 & 0.000902 & 125.796875 & 7.449219 \\
R2D\_2TFE\_TCN\_S & 50.069781 & 0.001429 & 0.001440 & 126.550781 & 2.191406 \\
R2D\_2TFE\_TCN\_M & 55.702710 & 0.001590 & 0.001601 & 127.515625 & 2.191406 \\
R2D\_2TFE\_TCN\_L & 65.021836 & 0.001856 & 0.001869 & 126.613281 & 2.195312 \\
R2D\_2TFE\_TCN\_XL & 86.963497 & 0.002482 & 0.002497 & 126.957031 & 2.339844 \\
R2D\_2TFE\_TRANSFORMER\_S & 41.083517 & 0.001172 & 0.001195 & 128.683594 & 2.535156 \\
R2D\_2TFE\_TRANSFORMER\_M & 55.094408 & 0.001572 & 0.001607 & 127.949219 & 2.550781 \\
R2D\_2TFE\_TRANSFORMER\_L & 63.903288 & 0.001824 & 0.001849 & 129.175781 & 2.964844 \\
R2D\_2TFE\_TRANSFORMER\_XL & 87.875360 & 0.002508 & 0.002531 & 129.437500 & 3.292969 \\
\bottomrule
\end{tabular}
\end{table*}

\section{Learning Scheme and Temporal Representation}
\label{sec:app_rl}

\begin{figure}[ht]
    \centering
    \includegraphics[width=0.48\textwidth]{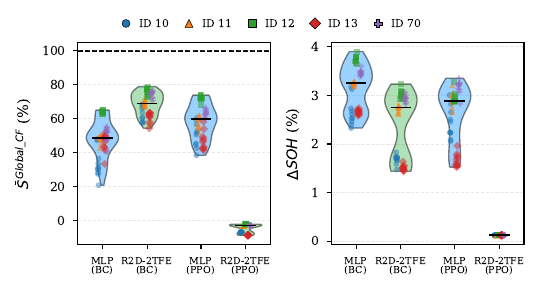}
    \caption{Normalized savings $\bar{S}^{Global\_CF}$ (left) and annual \gls{soh} loss (right) for the snapshot \gls{mlp} and the R2D\_2TFE encoder, each trained by behavior cloning and by \gls{ppo}. Violin distributions summarize ten seeds per site, dots indicate individual seeds coloured by site, and black bars denote means.}
    \label{fig:rl_bc}
\end{figure}

Throughout this paper, R2D denotes the proposed architecture trained offline by behavior cloning, and RL\_STD denotes the snapshot \gls{mlp} trained by \gls{ppo}; the two therefore differ in both the network structure and the training scheme. To separate these factors, this appendix crosses them: the snapshot \gls{mlp} and the R2D\_2TFE encoder are each trained once by behavior cloning from the 1D-MILP expert and once by \gls{ppo}. The two additional combinations serve this analysis only and do not change the definitions used in the main text. All four share the same simulation environment, the same test year and ten random seeds per site, and the behavior-cloned R2D and the snapshot \gls{ppo} policy are those reported in the benchmark study. Fig.~\ref{fig:rl_bc} shows the normalized savings and the annual \gls{soh} loss.

With the training scheme held fixed, the encoder accounts for the difference. Under behavior cloning, R2D\_2TFE reaches 68.5\% of the global optimum on average against 48.2\% for the snapshot \gls{mlp}, is better at every site, and shows both a smaller seed spread, at most 4.4~pp against up to 8.6~pp, and a lower annual \gls{soh} loss. Under \gls{ppo} the snapshot policy reaches 58.1\%, so the advantage of R2D over the \gls{rl} baseline does not follow from imitation alone.

Applying \gls{ppo} to the R2D encoder, by contrast, does not train. The policy converges to a near-idle battery at all five sites, with negative savings and an annual \gls{soh} loss of 0.12\%. The temporal observation enlarges the state space by orders of magnitude while the reward remains a delayed economic signal, and no useful policy is found within the same budget, whereas behavior cloning exploits the same encoder immediately. A staged scheme in which behavior cloning provides the initial policy and reinforcement learning refines it~\cite{yin2025boosting} was not evaluated for the temporal encoder here and requires additional design effort, but it is a direction worth investigating for both method families.

\end{document}